\documentclass[aps,pre,twocolumn,superscriptaddress,byrevtex,footinbib,bibnotes,longbibliography,bibnotes]{revtex4-1}
\usepackage{microtype}
\usepackage{amsmath}
\usepackage{amssymb}
\usepackage{graphicx}
\usepackage{color}
\definecolor{myblue}{rgb}{0.153,0.322,0.706}
\usepackage[colorlinks,linkcolor=myblue,urlcolor=myblue,citecolor=myblue]{hyperref}

\newcommand{\be}{\begin{equation}}
\newcommand{\ee}{\end{equation}}
\newcommand{\ra}{\rightarrow}

\newcommand{\reals}{\mathbb{R}}
\newcommand{\poi}{\text{Poisson}}
\newcommand{\mf}{\text{mf}}
\newcommand{\tPi}{\tilde\Pi}
\newcommand{\ER}{Erd\"os--R\'enyi~}
\newcommand{\IPR}{\text{IPR}}

\begin{document}

\title{Large deviations of random walks on random graphs}

\author{Francesco Coghi}
\email{francesco.coghi@gmail.com, f.coghi@qmul.ac.uk}
\affiliation{School of Mathematical Sciences, Queen Mary University of London, London E1 4NS, UK}

\author{Jules Morand}
\email{morandjules@gmail.com}
\affiliation{BioISI -- Biosystems \& Integrative Sciences Institute, Faculty of Sciences, University of Lisboa, Campo Grande C8, 1749-016 Lisboa, Portugal}

\author{Hugo Touchette}
\email{htouchette@sun.ac.za, htouchet@alum.mit.edu}
\affiliation{Department of Mathematical Sciences, Stellenbosch University, Stellenbosch 7600, South Africa}
\affiliation{National Institute for Theoretical Physics (NITheP), Stellenbosch 7600, South Africa}

\date{\today}

\begin{abstract}
We study the rare fluctuations or large deviations of time-integrated functionals or observables of an unbiased random walk evolving on \ER random graphs, and construct a modified, biased random walk that explains how these fluctuations arise in the long-time limit. Two observables are considered: the sum of the degrees visited by the random walk and the sum of their logarithm, related to the trajectory entropy. The modified random walk is used for both quantities to explain how sudden changes in degree fluctuations, similar to dynamical phase transitions, are related to localization transitions. For the second quantity, we also establish links between the large deviations of the trajectory entropy and the maximum entropy random walk. 
\end{abstract}

\maketitle

\section{Introduction}

Stochastic processes evolving on graphs are used to model a variety of natural and man-made phenomena, ranging from search algorithms and the spreading of infections, damages, attacks or rumors to the detection of communities in social networks and the efficient propagation of information in communication networks \cite{barrat2008,newman2010,pastor2015}. The focus in these applications is generally on averaged quantities such as mean first-passage times, commonly used as an efficiency measure in random search algorithms, or stationary distributions giving the average state occupation. Much less is understood about the occurrence of fluctuations or rare events far away from average or typical values, related, for example, to the rapid spreading of a disease in a dispersed population or faster-than-average location of a target node in a network.

How ``rare events'' are defined and how their probabilities are calculated depend on the application considered. One can study, for example, the evolution of $N$ random walkers on a graph and calculate the probability that $n$ of them reach a given node at a given time, with $n$ being much smaller or larger than the mean number of walkers expected on that node, as given by the stationary distribution of the random walkers. This type of ``occupation'' rare event was studied recently in \cite{kishore2011,kishore2012,kishore2013} and is relevant for investigating, for example, congestion in communication networks resulting from a high number of data packets reaching servers with limited capacity.

A different approach is to look at rare events arising from atypical initial conditions such as a low number of ``infected'' or ``damaged'' nodes evolving by contact or percolation dynamics to a large connected component \cite{altarelli2013,bianconi2017,bianconi2018,coghi2018,torrisi2018}. Similarly, one can study how an initially large population spread on a network becomes extinct in time \cite{dykman2008,lindley2014,hindes2016,hindes2017,hindes2017b} or how a process transitions, more generally, between macroscopically distinct states \cite{hindes2017c}. If the process is a deterministic dynamics, then a rare event arises in this case from the random choice of an initial condition propagated by the dynamics to a final random state.

In this paper, we follow a more dynamical approach initiated by De Bacco \emph{et al.}~\cite{bacco2015}, whereby a single random walker hops on a random graph and accumulates each time it jumps a certain ``cost'' related, for example, to the degree of the node visited or other characteristics of that node. If the random walk is ergodic, then the mean cost (i.e., the total cost divided by the total number of jumps) will converge with probability 1 to the ergodic average of the cost as the number of jumps increases, meaning that the cost of most trajectories of the random walk is very close to that average. However, for any large but finite number of jumps, there is a small probability that the mean cost will depart significantly from this average. Our goal is to calculate the probability of these rare cost fluctuations using large deviation theory \cite{dembo1998,hollander2000,touchette2009} and to understand how they arise dynamically from atypical trajectories of the random walk. This can be useful to target high or low connectivity regions in random graphs, so as to understand, for example, how congestion builds up in transport or data networks.

The cost that we consider is the mean degree of the nodes visited by an unbiased random walk on \ER random graphs, which was also considered in \cite{bacco2015}. Our contribution is to clarify some of the hypotheses used in that work to approximate the large deviation functions characterising the cost fluctuations, and to derive the so-called driven process, which is a modified random walk that explains how specific cost fluctuations are created in time \cite{jack2010b,chetrite2013,chetrite2014,chetrite2015}. This is useful, as we will see, to understand how dynamical phase transitions arising in cost fluctuations are linked to localization transitions, as first reported in \cite{bacco2015}. On a more practical level, the driven process can also be controlled to identify nodes with low or high degree, in addition to other graph properties, without knowing the detailed structure of the graph.

To complement these results, we consider another cost given by the sum of the logarithm of the degrees visited by the random walk, which is related to the trajectory entropy and entropy rate of the random walk. In this case, we re-obtain recent results related to the maximum entropy random walk \cite{burda2009,nechaev2017,gomez2008}, thereby providing a new interpretation of this random walk based on large deviation theory. We conclude by discussing the applicability of our results to other random processes, graph ensembles, and cost functions.

\section{Model and large deviations}
\label{secmod}

Let $G=(V,E)$ be an undirected graph with $V$ denoting the set of $N$ vertices and $E$ the set of $M$ edges. We consider a random walk $\{X_\ell\}_{\ell=1}^n$ evolving on this graph according to the transition probability
\be
\Pi_{ij} = \frac{A_{ij}}{k_i}
\label{eqprobtr1}
\ee
of going from node $X_\ell=i$ at time $\ell$ to node $X_{\ell+1}=j$ at time $\ell+1$. Here, $A_{ij}$ is the adjacency matrix of $G$
\be
A_{ij}=
\left\{
\begin{array}{lll}
1 & & i,j \text{ are connected}\\
0 & & \text{otherwise}
\end{array}
\right.
\ee
and $k_i$ is the degree of the starting node $i$, given by
\be
k_i =\sum_{j\in V} A_{ij}.
\ee
This choice of transition probability defines the \emph{unbiased random walk} (URW), as it has a uniform probability of going from node $i$ to any of its first neighbours $j\in \partial i$. 

The properties of the URW are well known \cite{barrat2008}. In particular, if $G$ has no disconnected component, then the random walk is ergodic and so has a unique stationary distribution, which is easily found to be proportional to the degree:
\be
p_i^* = \frac{k_i}{\sum_{i\in V} k_i} = \frac{k_i}{2M}.
\label{eqstatdist1}
\ee
The random walk is also reversible, as it satisfies the property of detailed balance with respect to the stationary distribution:
\be
p_i^* \Pi_{ij} = p_j^* \Pi_{ji}.
\ee
This can be used to transform $\Pi_{ij}$ to the symmetric matrix
\be
\hat\Pi_{ij}= (p_i^*)^{1/2} \Pi_{ij}  (p_j^*)^{-1/2},
\label{eqsym1}
\ee
which implies that the spectrum of $\Pi_{ij}$ is real \cite{bacco2015}.

Following the introduction, we assume that the random walk accumulates a cost in time given by
\be
C_n =\frac{1}{n} \sum_{\ell=1}^n f(X_\ell),
\label{eqobs1}
\ee
where $f$ is any function of the node state. This cost is also called a \emph{dynamical observable} in nonequilibrium statistical mechanics \cite{touchette2009}. Because of the ergodicity of the URW, we have that $C_n$ converges with probability 1 to the ergodic average
\be
\langle f(X)\rangle_{p^*}=\sum_{i\in V} p^*_i \, f(i)=: c^*
\label{eqobsave1}
\ee
in the long-time limit $n\ra\infty$. This concentration property of time averages, corresponding in mathematics to the ergodic theorem, is used in practice to estimate many properties of large graphs such as the degree distribution or more involved centrality measures, by running random walks on those graphs for long times. In particular, if we choose $f(X_\ell)=\delta_{X_\ell,i}$, then $C_n$ converges to the stationary probability $p_i^*$, whereas if $f(X_\ell)=k_{X_\ell}$, the degree visited by $X_\ell$, then $C_n$ converges to the average degree of the graph.

Here, we study the fluctuations of $C_n$ around the typical or concentration value $c^*$ by calculating its probability distribution $P_n(c)=P(C_n=c)$ in the limit of large $n$. Following the theory of large deviations \cite{dembo1998,hollander2000,touchette2009}, this distribution is known to have the exponentially decaying form
\be
P_n(c)= e^{-n I(c)+o(n)},
\ee
where $o(n)$ denotes corrections smaller than linear in $n$. We thus focus on studying the decay or \emph{rate function}, given by the limit
\be
I(c) = \lim_{n\ra\infty} -\frac{1}{n}\ln P_n(c),
\label{eqldt1}
\ee
which characterizes the fluctuations of $C_n$ to leading order in $n$. This function is positive, $I(c)\geq 0$, and is equal to zero for ergodic random walks only for $c^*$, so that $P_n(c)$ decays exponentially fast with the final time $n$, except at $c^*$ where it concentrates exponentially.

To obtain the rate function, we use the G\"artner--Ellis Theorem \cite{dembo1998,hollander2000,touchette2009}, which states that $I(c)$ is given by the Legendre transform of the \emph{scaled cumulant generating function} (SCGF)
\be
\Psi(s) = \lim_{n\ra\infty}\frac{1}{n}\ln \langle e^{nsC_n}\rangle,\qquad s\in\reals
\ee
if the latter is differentiable. For ergodic Markov chains, including ergodic random walks, the SCGF is known to be given by the logarithm of the dominant or principal eigenvalue $\zeta_{\max}$ of the following positive matrix \cite{dembo1998}:
\be
(\tPi_s)_{ij} = \Pi_{ij} e^{s f(i)},
\ee
called the \emph{tilted matrix}. Thus,
\be \label{SCGF_ergodic}
\Psi(s) = \ln \zeta_{\max}(\tPi_s).
\ee
Moreover, it is known that the principal eigenvalue is differentiable whenever the state space (here, the set of nodes) is finite \footnote{For finite, ergodic Markov chains, $\lambda(k)$ is actually analytic in $k$ \cite{dembo1998}. Care must be taken when considering the limit of infinite-size graphs, as well as graphs that are not all connected. This is discussed later in the text.}, so we can write in the end
\be 
I(c)= s_c c-\Psi(s_c),
\label{eqlf1}
\ee
where $s_c$ is the unique root of $\Psi'(s)=c$ \footnote{The unicity of the root follows from the fact that $\Psi(s)$ is convex by definition and strictly convex for ergodic processes on a finite state-space \cite{dembo1998}.}.

Obtaining the rate function is a difficult problem in general, as it is based on calculating the dominant eigenvalue of a positive matrix. This cannot be carried out analytically for most graphs, unless there are obvious structures or symmetries. However, the SCGF can be computed numerically fairly directly for any finite graph, since the tilted matrix $\tPi_s$ is typically sparse and can be transformed, as in \eqref{eqsym1}, to a symmetric matrix for which fast eigenvalue routines can be used. 

Following \cite{bacco2015}, we consider here \ER (ER) random graphs generated by connecting any two vertices with probability $p$, so that the probability of a graph $G$ having $N$ vertices and $M$ edges is the binomial distribution over the $N(N-1)/2$ possible edges:
\be
P_{N,p} (G)= p^{M}(1-p)^{N(N-1)/2-M}.
\ee
We also consider the sparse regime where the link probability is chosen as $p=\alpha/N$ with $\alpha>1$. In this case, it is known that the following properties apply in the ``thermodynamic limit'' where $N\ra\infty$ \cite[Chap.~12]{newman2010}:
\begin{enumerate}
\item The average degree $\langle k\rangle$, calculated over the whole ER graph ensemble, converges to $\alpha$.
\item Most graphs have a giant connected component of size proportional to $N$.
\item The empirical degree distribution $\hat P_{G}(k)$, giving the frequencies of the degrees in a given graph $G$, converges with probability 1 to the Poisson distribution,
\be
P_{\poi}(k)=\frac{\alpha^k e^{-\alpha}}{k!},
\ee
with average $\langle k\rangle=\alpha$.

\item The nodes become asymptotically independent as the probability $P(k'|k)$ of observing a node of degree $k'$ linked to a node of degree $k$ converges to
\be 
Q(k') =\frac{k'P_{\poi}(k')}{\langle k\rangle} = \frac{k'\alpha^{k'-1}e^{-\alpha}}{k'!},
\label{eqqdist1}
\ee
which does not depend on $k$.

\item There are degree-degree correlations in the giant component, which decay, however, with increasing $\alpha$ \cite{oles2011,tishby2018}.
\end{enumerate}

These properties are useful for deriving analytical approximations of $\Psi(s)$ and $I(c)$ for large ER graphs having a high mean connectivity $\alpha$, as found for the mean degree \cite{bacco2015}. Our goal here is to revisit these approximations, discuss their validity, and to consider a new observable related, as mentioned, to the maximum entropy random walk.

To understand how large deviations are created in time, we will also construct the driven process associated with a given fluctuation $C_n=c$. In our context, this process is a locally-biased version of the URW (see Appendix E of \cite{chetrite2014}) whose transition probability matrix is given by
\be
(\Pi_s)_{ij} = \frac{(\tPi_s)_{ij} r_s(j)}{r_s(i)e^{\Psi(s)}}= \frac{\Pi_{ij} e^{sf(i)}r_s(j)}{r_s(i)e^{\Psi(s)}},
\label{eqdriven1}
\ee
where $r_s$ is the eigenvector associated with the principal eigenvalue of $\tPi_s$ \footnote{This eigenvalue equation is different from the one considered in \cite{bacco2015} because of a different convention used for labeling the transition matrix elements, although the dominant eigenvalue giving the SCGF is the same.}:
\be
\tPi_s r_s = \zeta_{\max}(\tPi_s) r_s=e^{\Psi(s)} r_s.
\label{eqeig1}
\ee

The interpretation of the driven process, which is also called the \emph{auxiliary} or \emph{effective process} \cite{jack2010b}, follows what we mentioned in the introduction: it is the effective dynamics of the subset of paths of the random walk leading to a fluctuation $C_n=c$ away from the typical value $c^*$, and so it is, in that sense, the effective \emph{biased} random walk that explains how that fluctuation is created up to time $n$ \cite{chetrite2013,chetrite2014,chetrite2015}. To match the fluctuation $C_n=c$, the parameter $s$ must be chosen such that $\Psi'(s)=c$ or, equivalently, $I'(c) = s$. This is similar to fixing the equilibrium energy of the canonical ensemble by fixing its inverse temperature to the derivative of the entropy \cite{chetrite2013}. Here, we fix $s$ so as to transform an \emph{atypical} value or fluctuation $C_n=c$ seen for the original URW into a \emph{typical} value of $C_n$ for the biased random walk.

\section{Mean degree}

\begin{figure*}[t]
\centering
\begin{tabular}[b]{c}
\includegraphics[width=.3\linewidth]{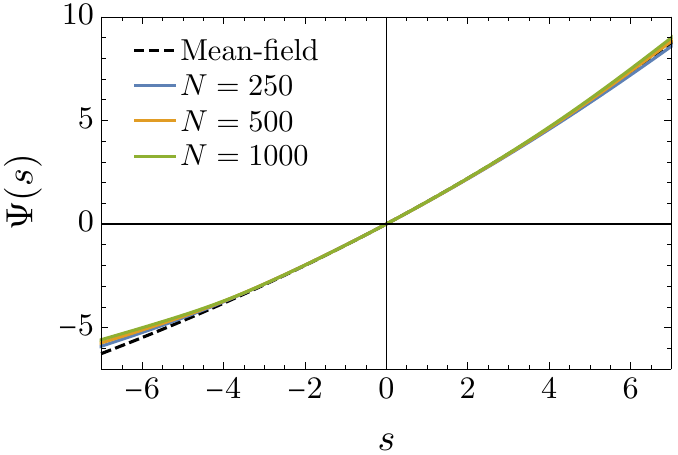}\\
\\
\includegraphics[width=.3\linewidth]{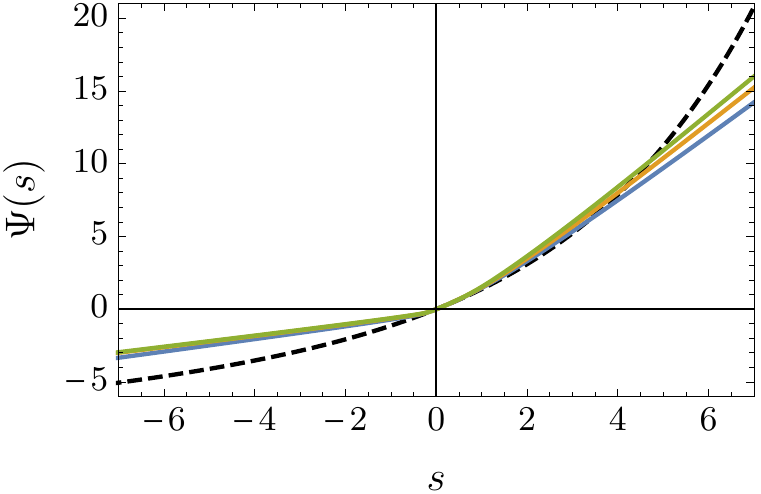}
\end{tabular}%
\hspace*{0.2in}%
\begin{tabular}[b]{c}
\includegraphics[width=.3\linewidth]{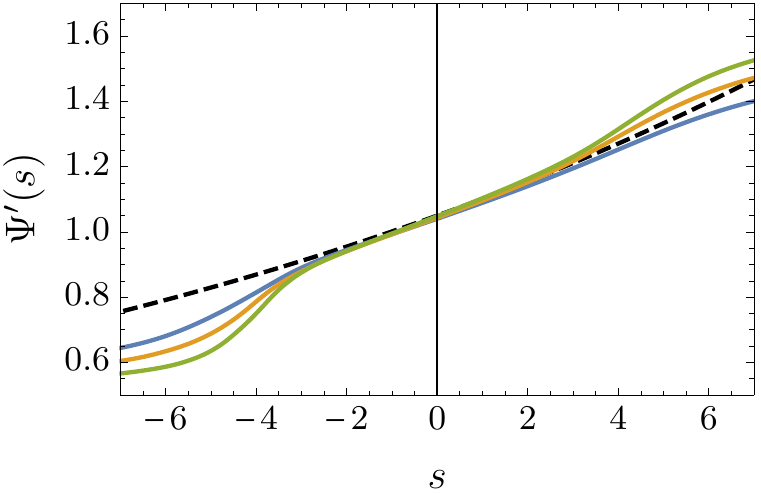}\\
\\
\includegraphics[width=.3\linewidth]{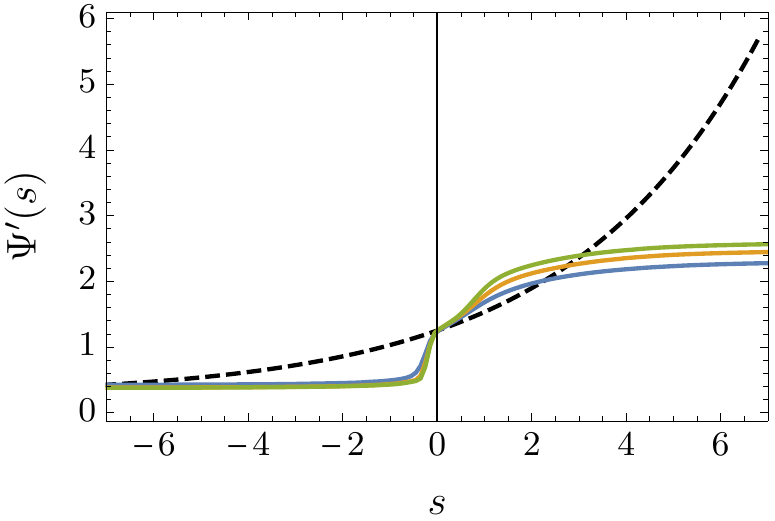}
\end{tabular}%
\hspace*{0.2in}%
\begin{tabular}[b]{c}
\includegraphics[width=.3\linewidth]{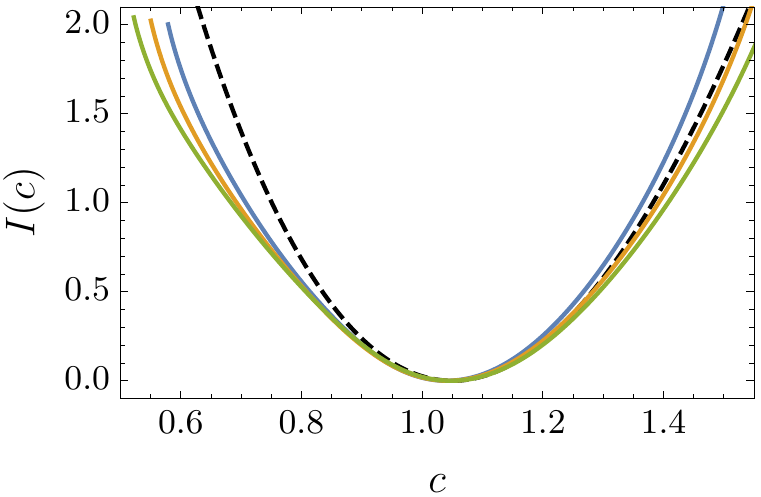}\\
\\
\includegraphics[width=.3\linewidth]{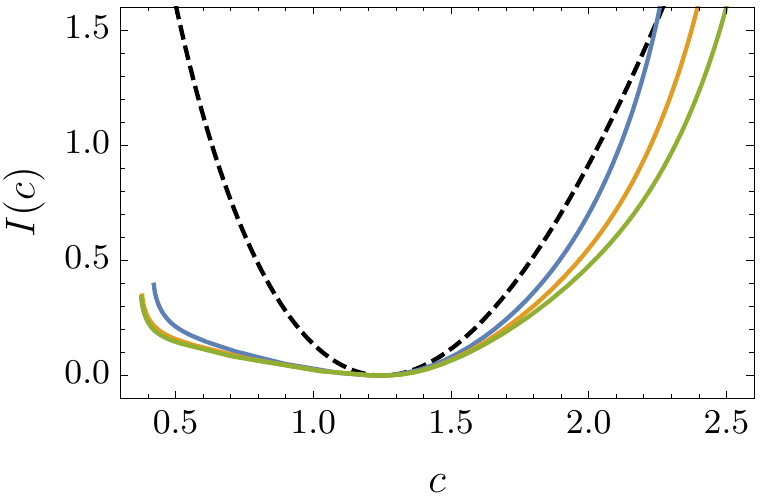}
\end{tabular}
\caption{(Color online) Large deviation functions of the mean degree for the URW on ER graphs with $\alpha=20$ (top row) and $\alpha=4$ (bottom row). Left column: Scaled cumulant generating function $\Psi(s)$. Middle column: Derivative of $\Psi(s)$. Right column: Rate function $I(c)$ obtained by Legendre transform of $\Psi(s)$. All results were obtained by averaging over 100 graphs.}
\label{fig1}
\end{figure*}

The first observable or cost that we consider is the mean degree, defined by
\be 
C_n = \frac{1}{n \alpha}\sum_{\ell=1}^n k_{X_\ell},
\label{eqmeandeg1}
\ee
where $X_\ell$ is the node visited by the URW at time $\ell$. This observable, with $\alpha$ in the denominator, is normalized to
\be
\langle C_n\rangle=\sum_{i=1}^N p_i^* \frac{k_i}{\alpha} = \sum_{k\geq 0} P_{\poi} (k) \frac{k}{\alpha}=1
\ee if we average the random walk in an annealed way over the whole ensemble of ER graphs. In our case, we consider random instances of the ER ensemble (quenched regime) and initialise the URW on the giant component, so as to avoid non-ergodic effects related to small disconnected components of the graph or single disconnected nodes \cite{bacco2015}. As a result, the stationary average with $P_\poi$ above must be computed instead with $Q(k)$, as given by \eqref{eqqdist1}, since we are considering nodes that are necessarily connected $(k\geq 1$), leading to
\be
\langle C_n\rangle_Q = \sum_{k>0} Q(k) \frac{k}{\alpha} = \frac{\alpha+1}{\alpha}=c^*
\ee
for the typical or concentration value of $C_n$. This applies in the sparse regime if $N$ and $\alpha$ are large enough, in which case the giant component is representative of the ``bulk'' properties of most graphs in the ER ensemble \cite{tishby2018}. This point is important; we will come back to it to understand the typical and large deviation properties of the URW.

\subsection{Large deviations}

To describe the fluctuations of $C_n$ around $c^*$, we calculated the SCGF by finding numerically the largest eigenvalue of the tilted matrix $\tPi$, symmetrized according to \eqref{eqsym1}, and by averaging over 100 graphs to obtain results that are representative of the ER ensemble \footnote{The averaging over graphs does not change the fact that $\Psi(s)$ is differentiable for finite graphs.}. The results are shown for different graph sizes in the left column of Fig.~\ref{fig1} for $\alpha =20$ (top) and $\alpha=4$ (bottom), and are also compared with a random degree or \emph{mean-field} (mf) approximation of $\Psi(s)$ proposed in \cite{bacco2015}, having the form
\be
\Psi_{\mf}(s) = \ln \langle e^{s k/\alpha}\rangle_Q = \alpha e^{s/\alpha}+\frac{s}{\alpha} -\alpha.
\label{eqmf1}
\ee
The results for $\alpha=20$ do not show, as can be seen, much variation across different graph sizes, and agree relatively well with the mean-field approximation and its derivative, shown in the middle column, which reproduces the mean
\be
\Psi_\mf'(0)=\frac{\alpha+1}{\alpha}.
\ee 
By comparison, the results obtained for $\alpha=4$ show more variations for each system size, even though the results are averaged over graphs, and are not well reproduced by the mean-field approximation. The mean $c^*$ is correctly reproduced, but the derivative of $\Psi(s)$ varies rapidly around $s=0$ and saturates quickly for $|s|\gg 1$ compared to $\Psi_\mf'(s)$, which is a shifted exponential. 

The same can be noted for the rate function, obtained by computing the Legendre transform of $\Psi(s)$ according to \eqref{eqlf1}. For $\alpha=20$, the result agrees well with the mean-field rate function, given by the Legendre transform of $\Psi_\mf(s)$:
\be
I_\mf(c) =1+\alpha -\alpha c+\alpha\left(c-\frac{1}{\alpha}\right)\ln \left(c-\frac{1}{\alpha}\right)
\ee
for $c\geq 1/\alpha$. This function is parabolic around $c^*$ and scales like $\alpha c\ln c-\alpha c$ for $c\gg c^*$, predicting that $P_n(c)$ concentrates in a Gaussian way around $c^*$, with a variance given by \cite{touchette2009}
\be
\frac{\Psi''_\mf(0)}{n}= \frac{1}{nI''_\mf(c^*)}=\frac{1}{\alpha n},
\ee
and that its right tail decays according to
\be
P_n(c)\sim c^{-n \alpha c} e^{n\alpha c},\qquad c\gg c^*.
\ee

This applies for $\alpha=20$ and for highly connected graphs, in general, having a large average degree. For $\alpha=4$, the rate function departs from the mean-field approximation: the former lies under the latter, implying that fluctuations of $C_n$ in low-connectivity graphs are more likely than predicted by the mean-field approximation. This is especially true for $c<c^*$. There we see that $I(c)$ is close to $0$ and increases nearly linearly as $c\ra c_{\min}=3/8$ because of the rapid increase of $\Psi'(s)$ left of $s=0$ \footnote{We say ``nearly linearly'' because $\Psi'(s)$ does not have a jump singularity when $N<\infty$, which is a necessary condition for $I(c)$ or its convex envelope to have a linear part \cite{touchette2009}.}. Such a linear part in the rate function seems to signal the appearance of a dynamical phase transition (DPT), interpreted as a ``co-existence'' of random paths that visit nodes with low degree and paths that visit the whole graph.

These results and observations were more or less already noted in \cite{bacco2015}. To understand them, note that $C_n$ is the sum of the degrees visited by the random walk in time, and does not carry, as such, all the information about that walk. In fact, we know that the node degrees are asymptotically uncorrelated for ER graphs, as noted before, which means that $C_n$ is essentially a sum of independent degrees -- the random degrees visited in time -- which are identically distributed, in the limit of large graphs, according to $Q(k)$. This explains the mean-field result: for sample means of independent and identically distributed (iid) random variables, the SCGF reduces to the cumulant of one random variable with distribution $Q$, leading to \eqref{eqmf1}. 

This explanation of the mean-field approximation is different from the one given in \cite{bacco2015}. There it is claimed that the eigenvector $r_s(i)$ depends on the node $i$ only via its degree $k_i$ and that this property leads, in the spectral equation \eqref{eqeig1}, to the mean-field expression of the SCGF. Numerically, we do not find that $r_s(i)$ is a function of the degree $k_i$ either exactly or asymptotically for large graphs \footnote{To test this, we have calculated the dispersion (standard error over mean) associated with the distribution of $r_s(i)$ over all nodes $i$ having the same degree, to find that it does not decay with $N$. This was verified for different degrees, as well as different values of $\alpha$ and $s$.}. Moreover, it can be checked that the spectral equation \eqref{eqeig1} does not yield the cumulant of $Q$ if $r_s(i)$ is assumed to be a function of the degree -- the resulting equation is similar to \eqref{eqmf1}, but has an extra degree term that cannot be eliminated to obtain that cumulant \footnote{That term should also appear as an extra $k$ term in Eq.~(11) of \cite{bacco2015}.}.

To properly understand the mean-field approximation, note that $C_n$ has the form of a sum of random variables $k_1,k_2,\ldots,k_n$ forming, technically, the \emph{visible layer} of a \emph{hidden Markov chain}, defined by
\be
k_\ell = f(X_\ell) = k_{X_\ell},
\ee 
where $X_1,X_2,\ldots,X_n$ is the Markov chain (i.e, the URW) in the \emph{hidden layer}. In general, the evolution of the visible layer is not Markovian, especially if it is a deterministic ``coarse-graining'' of the hidden layer \cite{ephraim2002}, as is the case here (coarse-graining from node to degree). The URW on ER graphs is special, in that the visible layer happens to be iid because of the uncorrelated nature of the giant component when $N\ra\infty$ and $\alpha$ is large enough. 

With this explanation, it should be clear that there are not one but two assumptions involved in the mean-field approximation: 1- the degrees visited are uncorrelated, and 2- the frequency of the degrees visited by the random walk is $Q$. These assumptions play at different levels depending on the value of $N$ and $\alpha$. For $\alpha=20$, we have found that replacing $Q$ in the expectation defining $\Psi_\mf(s)$ by the actual degree distribution $\hat P_{G}(k)$ of the graph $G$ considered gives a better approximation to $\Psi(s)$, which shows that the uncorrelated assumption is verified, but that the graphs considered were not large enough to have $\hat P_{G}=Q$. The different $\Psi(s)$ obtained for different graphs are then accounted for by the fluctuations of $\hat P_{G}$ around $Q$, which can be described in principle by large deviation theory using Sanov's Theorem \cite{hollander2000}. 

For $\alpha=4$, however, the mean-field approximation is not good even when calculated with $\hat P_{G}$ because the uncorrelated assumption is not verified: there are degree-degree correlations in the different parts of the graph (center or edge) visited by the URW, which get stronger as $|s|\ra \infty$. Moreover, in this case there is a cut-off in the maximum degree seen in finite-size graphs, compared with the mean-field prediction, leading to the saturation of $\Psi'(s)$ seen for $s\ra \infty$. This saturation or ``linearization'' effect arises whenever the quantities sampled (here the degrees) are artificially bounded \cite{rohwer2014}. Finally, note that the mean-field approximation wrongly predicts that the minimum value of $C_n$ is $1/\alpha$ because the degrees visited can be equal to 1 at all times assuming they are iid. For the URW on the giant component, a degree 1 can only be followed by a degree of at least 2, leading to $c_{\min} = 3/(2\alpha)$.

\subsection{Biased random walk}

\begin{figure*}[p]
\centering
\includegraphics{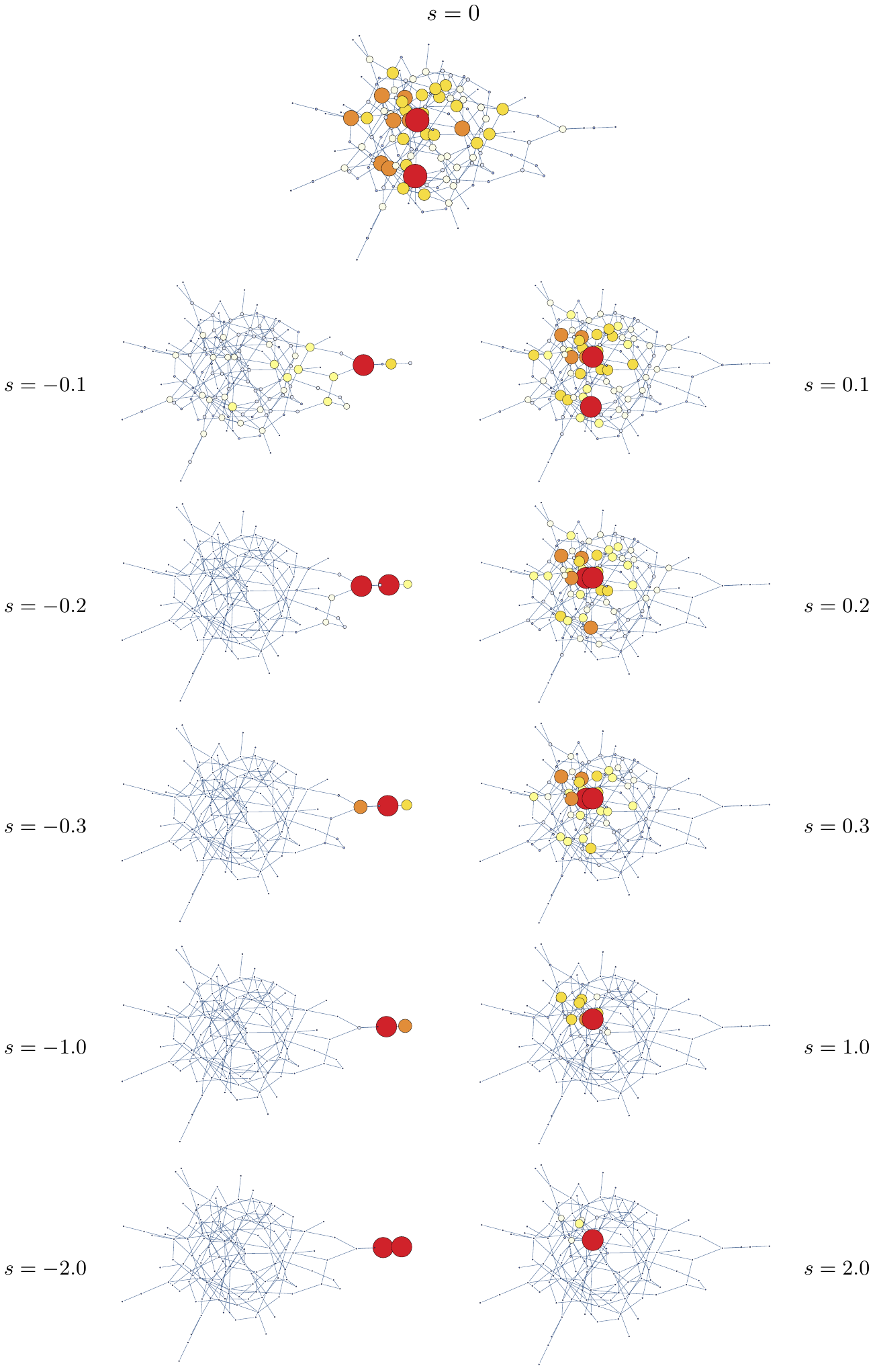}
\caption{(Color online) Graphical illustration of the stationary distribution $p_s$ of the BRW on an ER graph with 250 nodes. The same graph is plotted for different values of $s$ by sizing each node $i\in V$ proportionally to $p_s(i)$.}
\label{fig2}
\end{figure*}

To understand how fluctuations of the mean degree arise dynamically, we now study the driven process, which is the biased random walk (BRW) with transition matrix $\Pi_s$ defined in \eqref{eqdriven1}. To this end, we calculate its stationary distribution, which is known \cite{chetrite2014} to be given by 
\be
p_s(i) = r_s(i) l_s(i),\qquad i\in V,
\label{eqsbrw1}
\ee
where $r_s$ is the ``right'' eigenvector of the tilted matrix $\tPi_s$ satisfying \eqref{eqeig1}, whereas $l_s$ is its ``left'' eigenvector satisfying 
\be
l_s \tPi_s =e^{\Psi(s)} l_s
\ee
as a row vector. These are normalized so that
\be
\sum_{i\in V} l_s(i)=1,\qquad \sum_{i\in V} r_s(i)l_s(i)=1.
\label{eqnorm1}
\ee

The results are presented in Fig.~\ref{fig2}, which shows, for different values of $s$, the random graph considered in our computation with each node $i\in V$ sized proportionally to the value of $p_s(i)$. The results obtained are specific to that random graph, but are generic in the way that $p_s(i)$ concentrates on different nodes of the graph as a function of the tuning parameter $s$. For $s=0$, in particular, $p_0$ is the stationary distribution $p^*$ of the URW, which means that the nodes' sizes seen in the top graph of Fig.~\ref{fig2} are proportional to their degree, as in \eqref{eqstatdist1}.

\begin{figure}[t]
\centering
\begin{tabular}[b]{c}
\includegraphics[width=2.7in]{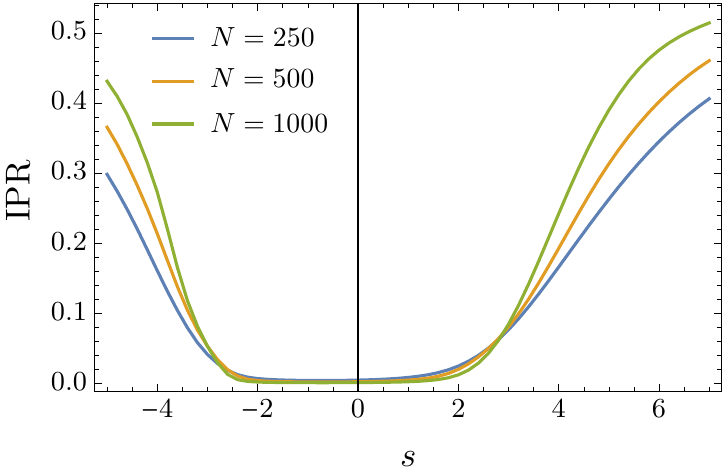}\\
\\
\includegraphics[width=2.7in]{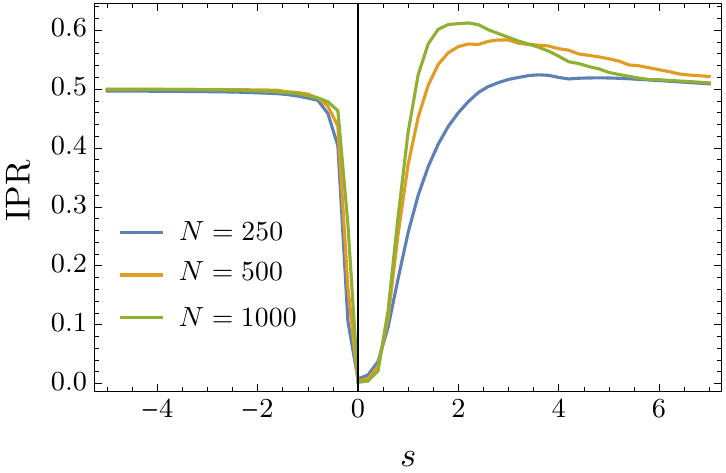}
\end{tabular}
\caption{(Color online) Inverse participation ratio (IPR) of the biased stationary distribution $p_s$ for the mean degree. Top: $\alpha=20$. Bottom: $\alpha=4$.}
\label{fig3}
\end{figure}

As we increase $s$, we see that $p_s$ concentrates on nodes with higher degree, as the URW is deformed by $\Pi_s$ to a BRW that visits higher-degree nodes in order to achieve a typical cost $c=\Psi'(s)$ larger than the mean cost $c^*$ achieved by the URW. In the limit $s\ra\infty$, the BRW reaches the largest cost, realized not by the largest-degree node, since the random walk has no self-loops or self-jumps, but by the two connected nodes with largest degrees. These nodes are likely to be unique and isolated, as noted in \cite{bacco2015}, since ER graphs have few high-degree nodes, so the BRW typically concentrates, for large $s$, to a very small ``island'' of highly-connected nodes, located in Fig.~\ref{fig2} inside the graph because of the drawing algorithm used. The degree correlations in this island are necessarily different from those found in the whole ER ensemble and the giant component, which explains why the mean-field approximation is not good for large $s$.

By decreasing $s$ from 0, we obtain a different picture. Then the BRW moves to low-degree nodes, located on the ``edge'' of the graph, so as to accumulate in the long-time limit a cost $c=\Psi'(s)$ lower than the mean $c^*$. As $s\ra -\infty$, it further concentrates on the lowest-degree nodes, reaching first tree-like nodes, as seen in Fig.~\ref{fig2}, and ultimately dangling pairs of nodes or ``hairs'' with degrees $k=1$ and $k=2$, which also leads to a strong departure from the mean-field assumption that all degrees be accessible from any node.

These transitions to highly and lowly connected nodes are clearly seen in Fig.~\ref{fig3} in the plot of the \emph{inverse participation ratio} (IPR) \footnote{De Bacco \emph{et al.}~\cite{bacco2015} define the IPR differently with what corresponds in our notation to the right eigenvector $r_s(i)$ raised to the power 4. Here, we follow the definition used in quantum mechanics, noting that $r_s(i)$ and $r_s(i)^2$ have no probabilistic interpretation.}, defined as
\be
\IPR =\sum_{i\in V} p_s(i)^2.
\ee
There we see that the IPR is close to $0$ when $s=0$, since $p_0=p^*$ is spread over all the nodes, and grows toward $1/2$ as $s\ra\pm\infty$, since $p_s$ then concentrates on two nodes in both limits. The localization is quicker for $s<0$, especially if we consider low connectivity graphs ($\alpha=4$), because the degree distribution is then less spread and more skewed: the mean degree is closer to the lowest than the largest degree, so localization toward hairs is faster than toward high-degree nodes. 

The transition to the hairs also explains why $\Psi'(s)$ is steep below $s=0$, leading to the near-linear part of $I(c)$. One way to interpolate between the lowest and the mean degrees is for the random walk to reach one hair and stay there for some fraction of the total time $n$, before moving to the bulk of the graph for the remaining fraction of time. This realizes the ``co-existence'' of paths mentioned before and predicts, by analogy with Markov chains with absorbing states (see Appendix~\ref{appsec}), a linear section of $I(c)$ with a slope equal to the decay probability of the random walk from hair to bulk. In this sense, the transition is not a switching between different fluctuation mechanisms, as seen, for example, in \cite{tsobgni2016}, but the result of an absorbing-like dynamics where the random walk survives in hairs for a given fraction of the total time. This interpretation is consistent with the results of \cite{bacco2015}, showing that the gap between the first two largest eigenvalues of the tilted matrix $\tPi_s$ decreases as $N\ra\infty$ around the transition (see their Fig.~5). Such a closing of the gap is also seen in Markov chains that become absorbing in some limit.

Whether this transition is a ``genuine'' DPT that becomes sharp in the thermodynamic limit is not clear at this point. For $N<\infty$, the transition is rounded or smeared \cite{dufresne2018} because the URW has a finite probability  to reach hairs from the bulk, so it is not strictly absorbing (see Appendix~\ref{appsec}). For the transition to become sharp, one needs to show that this return probability vanishes as $N\ra\infty$, which happens if the bulk is a complete graph of size $N$ or, more generally, if the degrees of the nodes in the bulk grow uniformly in $N$, none of which applies to ER graphs in the sparse limit. Nevertheless, the URW can become trapped in the bulk as a result of more nodes being accessible in the thermodynamic limit. More numerical or analytic work is required to verify this and to determine, more precisely, the scaling of the return probability with $N$.

If confirmed, the DPT must be first-order and not second-order, as claimed in \cite{bacco2015}, since it is associated with a jump in the derivative of the SCGF, which is rounded off again for finite-size graphs. Moreover, the DPT cannot be accounted for by the mean-field approximation or by assuming that the degrees visited by the URW form a Markov chain, as both approaches predict a smooth SCGF. Degree-degree correlations must be involved, since the BRW again concentrates as a function of $s$ on a restricted set of nodes having different degree-degree correlations compared to the giant component or the whole ER ensemble \cite{tishby2018}.

\section{Mean entropy}

\begin{figure*}[t]
\centering
\begin{tabular}[b]{c}
\includegraphics[width=.3\linewidth]{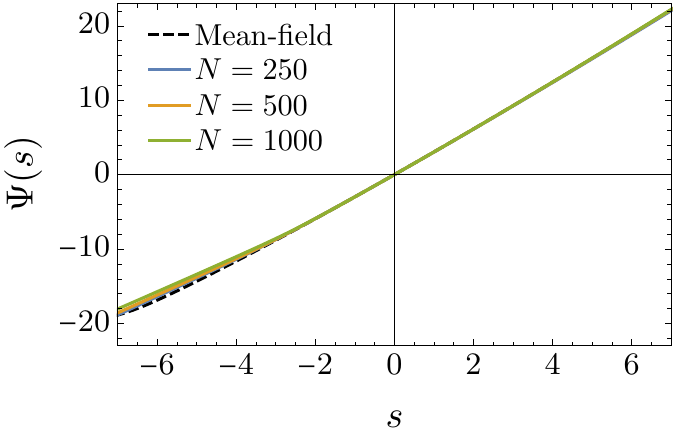}\\
\\
\includegraphics[width=.3\linewidth]{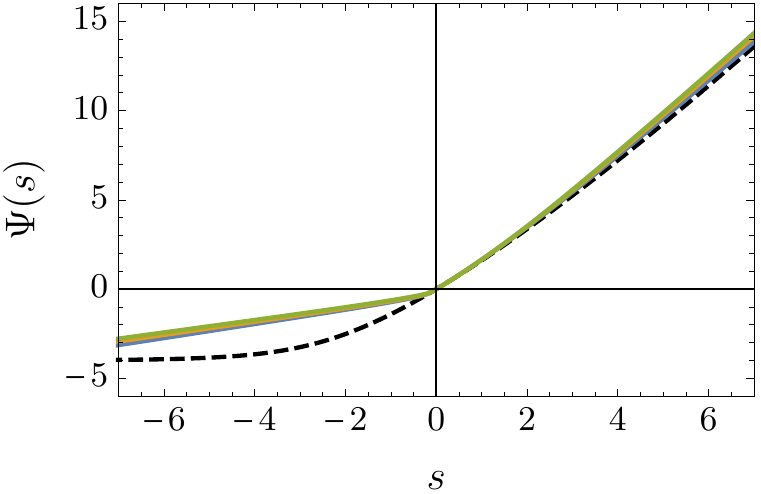}
\end{tabular}%
\hspace*{0.2in}%
\begin{tabular}[b]{c}
\includegraphics[width=.3\linewidth]{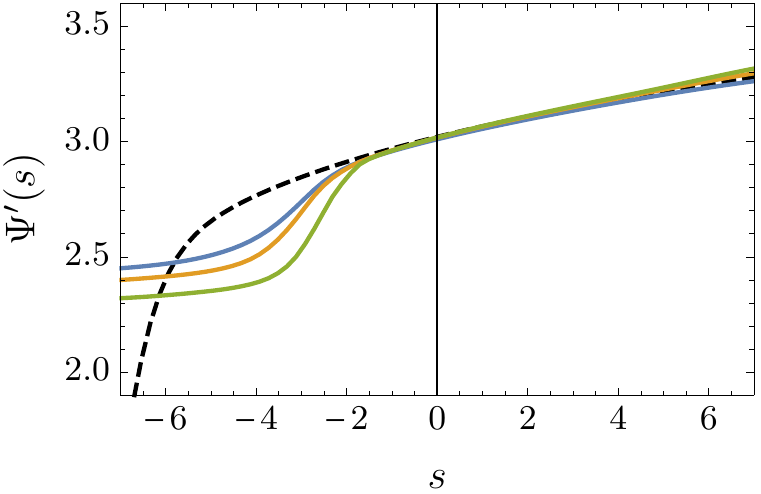}\\
\\
\includegraphics[width=.3\linewidth]{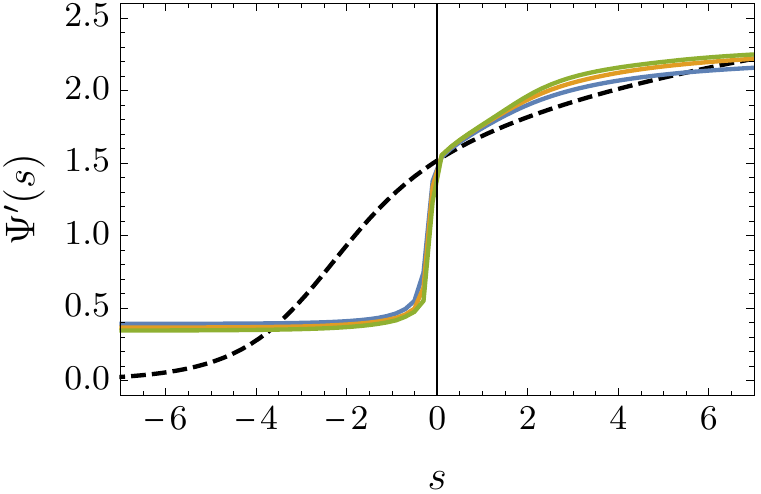}
\end{tabular}%
\hspace*{0.2in}%
\begin{tabular}[b]{c}
\includegraphics[width=.3\linewidth]{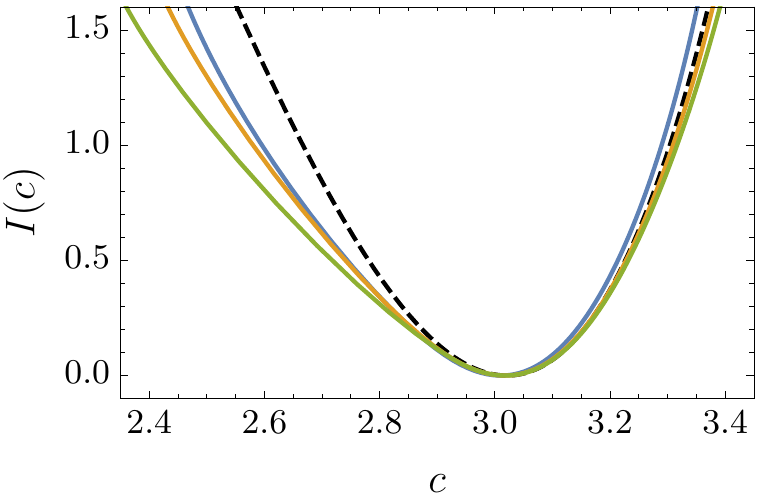}\\
\\
\includegraphics[width=.3\linewidth]{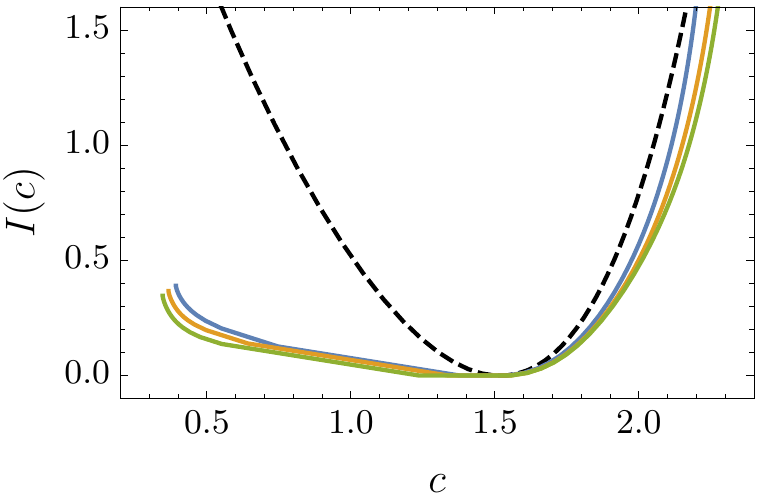}
\end{tabular}
\caption{(Color online) Large deviation functions of the mean entropy for the URW on ER graphs with $\alpha=20$ (top row) and $\alpha=4$ (bottom row). Left column: Scaled cumulant generating function $\Psi(s)$. Middle column: Derivative of $\Psi(s)$. Right column: Rate function $I(c)$ obtained by Legendre transform of $\Psi(s)$.}
\label{fig4}
\end{figure*}

As a second observable, we consider
\be
C_n = \frac{1}{n}\sum_{\ell=1}^n \ln k_{X_\ell},
\ee
where $k_{X_\ell}$ is, as before, the degree visited by the random walk at time $\ell$. This variation of the mean degree, involving the logarithm, is related to the entropy rate of the URW \cite{cover1991,gomez2008,sinatra2011}, defined by the limit
\be
h=\lim_{n\ra\infty} \frac{1}{n}H(X_1,\ldots,X_n),
\label{eqentr1}
\ee
where $H(X_1,\ldots, X_n)$ is the entropy of the probability distribution $P(x_1,\ldots,x_n)$ over the possible paths $x_1,\ldots,x_n$ of the URW. For ergodic random walks, it is known \cite{cover1991} that the limit reduces to
\be
h=-\sum_{i,j} p^*(i) \, \Pi_{ij} \ln \Pi_{ij},
\ee
where $p^*$ is the stationary distribution of the random walk and $\Pi_{ij}$ is its transition probability, given for the URW by \eqref{eqprobtr1}. As a result, we find 
\be
P(x_1,\ldots,x_n)=P(x_1)\prod_{\ell=1}^{n-1} k^{-1}_{x_\ell},
\ee
for the connected paths, where $P(x_1)$ is the initial distribution of the URW. Up to boundary terms at $\ell =1$ and $\ell=n$ that do not influence the large deviations, we can therefore write
\be
C_n = -\frac{1}{n}\ln P(X_1,\ldots,X_n),
\ee
leading to
\be
h=\lim_{n\ra\infty} \langle C_n\rangle= \langle\ln k_X\rangle_{p^*}=\frac{1}{2M} \sum_{i\in V} k_i \ln k_i.
\ee
The entropy rate is also called the \emph{Kolmogorov--Sinai entropy} and represents, from \eqref{eqentr1}, the mean information per step generated by the URW  \cite{cover1991}. The observable  $C_n$, on the other hand, is sometimes called the fluctuating trajectory entropy \cite{seifert2005} or the \emph{self-process} \cite{touchette2009}, since it is a random variable of the process (the random walk) involving the very distribution of that process.

\subsection{Large deviations}

As before, we are interested in the fluctuations of $C_n$ around its mean, corresponding, as above, to the entropy rate. The tilted matrix now has the form
\be
\tPi_{ij} = \Pi_{ij} e^{s \ln k_i} = \Pi_{ij} k_i^s=A_{ij} k_i^{s-1}.
\label{eqtmme1}
\ee
The eigenvalue equation \eqref{eqeig1} still cannot be solved exactly and so we resort, as for the mean degree, to exact numerical diagonalization to find the largest eigenvalue giving the SCGF and, by Legendre transform, the rate function. 

The results are presented in Fig.~\ref{fig4}. They are similar to those obtained for the mean degree, in that we also find a relatively good agreement between the exact results and the mean-field approximation for highly connected graphs ($\alpha=20$), but not for low connected graphs ($\alpha=4$), although the saturation effect for $s<0$ is present in both. The minimum value $C_n=0$ predicted by the mean-field approximation is also wrong in both cases, being $c_{\min}=\ln(2)/2$ in the URW. Note that the mean-field SCGF is now given by
\be
\Psi_\mf(s) = \ln \langle k^s\rangle_Q.
\ee 
Contrary to the mean degree, the expectation with $Q$ has no closed-form expression, so we compute it numerically by truncating the sum to a high degree. From the SCGF, we then compute the Legendre transform numerically to obtain the mean-field rate function shown in the right plots of Fig.~\ref{fig4}.

\begin{figure}[t]
\centering
\begin{tabular}[b]{c}
\includegraphics[width=2.7in]{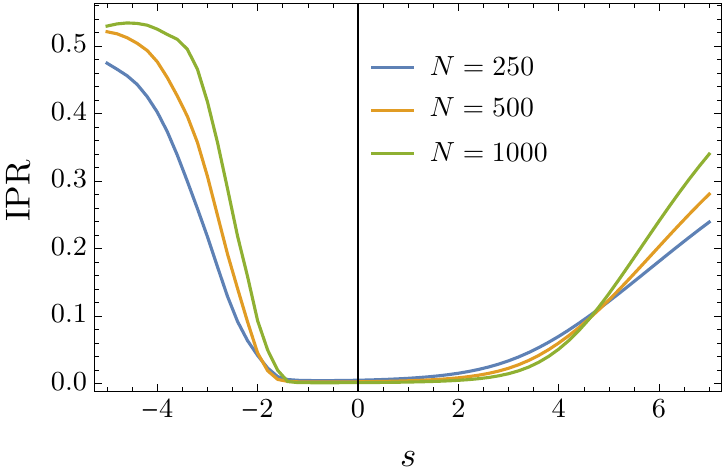}\\
\\
\includegraphics[width=2.7in]{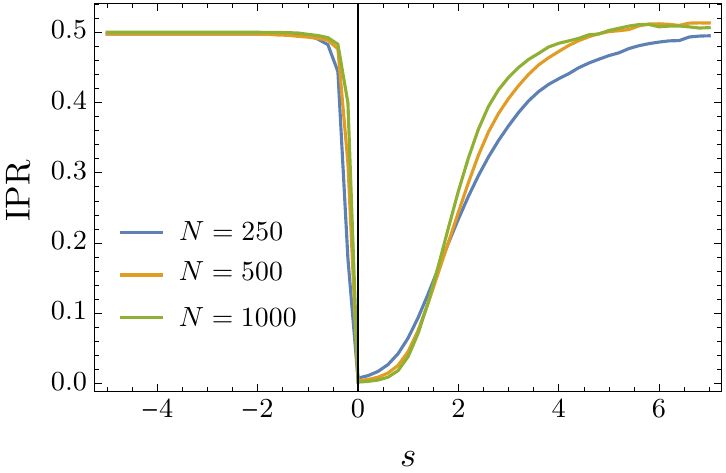}
\end{tabular}
\caption{(Color online) Inverse participation ratio (IPR) of the biased stationary distribution $p_s$ for the mean entropy. Top: $\alpha=20$. Bottom: $\alpha=4$.}
\label{fig5}
\end{figure}

Looking at the results for $\alpha=4$, we can see that there is an abrupt transition in $\Psi'(s)$, leading as before to a near-linear part in the rate function $I(c)$. The interpretation of this rounded transition is the same as for the mean degree: it comes from the survival probability of the random walk staying in a hair for a fraction of the total time $n$, and is more abrupt than for the mean entropy because the logarithm has the effect of lowering the value of $C_n$. The power-law form in $k$ of the mean-field SCGF, coming also from the logarithm, is less affected by linearization effects, explaining why the agreement for $s$ large is good even for $\alpha$ small. This is confirmed by the plots of the IPR which show a slower localization for $s>0$ but faster localization for $s<0$, compared with the mean degree (Fig.~\ref{fig5}). We do not show the graph representation of the stationary distribution $p_s$ of the BRW, as it is similar to the one found for the mean degree.


\subsection{Maximum entropy random walk}

The form of the tilted matrix for the mean entropy leads to an interesting relation between the BRW defined by \eqref{eqdriven1} and the \emph{maximum entropy random walk} (MERW) defined as the random walk on $G$ achieving the largest entropy rate \cite{gomez2008} or, equivalently, the random walk that has a uniform probability distribution for all paths connecting any two nodes \cite{burda2009}. For $s=1$, we indeed find from \eqref{eqtmme1} that the tilted matrix $\tPi$ is the adjacency matrix, which means that the transition matrix of the BRW is
\be
(\Pi_1)_{ij} = \frac{A_{ij}}{e^{\Psi(1)}}\frac{r_1(j)}{r_1(i)},
\ee
where $r_1$ is the eigenvector of $A$ and $e^{\Psi(1)}$ is the dominant eigenvalue of $A$. This reproduces the known transition matrix of the MERW \cite{burda2009}, which is expected since the BRW corresponds to the URW conditioned on the value of $C_n$. As we condition here on the entropy rate, the URW conditioned on reaching the maximum entropy rate must coincide with the MERW. 

\begin{figure}[t]
\centering
\begin{tabular}[b]{c}
\includegraphics[width=2.7in]{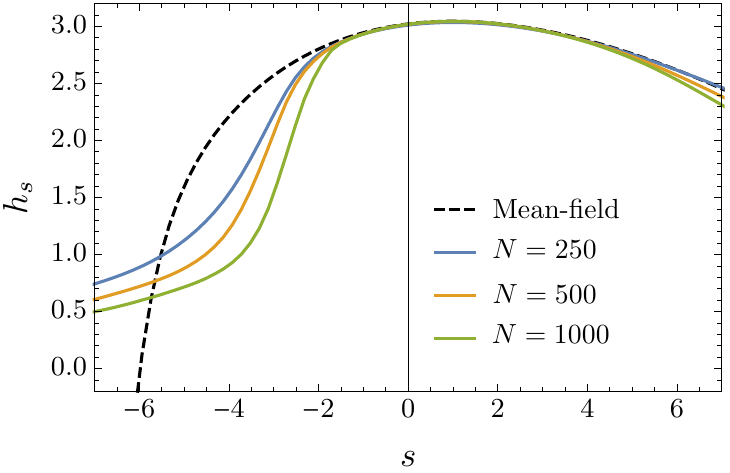}\\
\\
\includegraphics[width=2.7in]{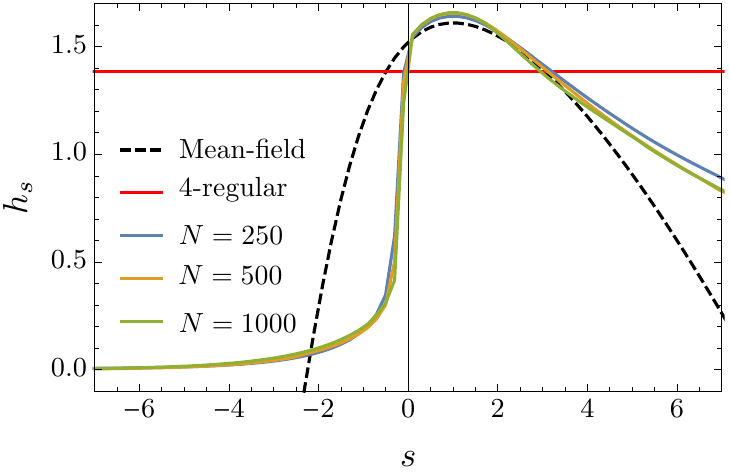}
\end{tabular}
\caption{(Color online) Entropy production $h_s$ of the BRW as a function of the tilting parameter $s$. Top: $\alpha=20$. Bottom: $\alpha=4$.}
\label{fig6}
\end{figure}

This result is confirmed by calculating the entropy rate of the BRW:
\be
h_s = -\sum_{i,j\in E} p_s(i) (\Pi_s)_{ij} \ln\, (\Pi_s)_{ij},
\ee
which can be expressed in terms of the SCGF as
\be
h_s = \Psi(s) +(1-s)\Psi'(s).
\ee
This follows by substituting in $h_s$ the expressions of $p_s$ and $\Pi_s$ for the BRW and using the normalization condition \eqref{eqnorm1}. It can be checked that this expression has a maximum at $s=1$, which is a global maximum because $h_1$ represents the maximum entropy rate. This can be seen in Fig.~\ref{fig6}, which shows $h_s$ as a function of $s$ for graphs with $\alpha=20$ (top) and $\alpha=4$ (bottom). The value at $s=1$ corresponds to the entropy rate of the MERW,
\be
h_1 = \Psi(1) = \ln \zeta_{\max}(A),
\ee
while
\be
h_0 = \Psi(0) +\Psi'(0)=\Psi'(0) = h
\ee
is the entropy rate of the URW. For comparison, we also show in Fig.~\ref{fig6} the entropy rate obtained from the mean-field approximation, as well as the entropy rate of the BRW on an $\alpha$-regular graph, which is constant, since $C_n$ can only take the value $\ln\alpha$, so that $\Psi(s)=s \ln\alpha$ and, therefore, $h_s=\ln \alpha$. Finally, note that $h\ra 0$ as $s\ra \infty$ or $s\ra-\infty$, since the BRW gets localized in both limits onto two sites, on which it oscillates in a deterministic way.

\section{Conclusion}

We have shown in this paper how to use tools from large deviation theory to study the fluctuations of time-integrated costs or observables of random walks evolving on random graphs, and how these fluctuations can be understood in terms of modified or biased random walks that represent, in the long-time limit, a random walk conditioned on reaching a certain fluctuation. Our results clarify the source and nature of apparent fluctuation transitions, first reported in \cite{bacco2015}, by proposing a mechanism explaining them, and provide insights about the maximum entropy random walk \cite{burda2009,nechaev2017,gomez2008}, which is equivalent to an unbiased random walk conditioned on reaching the maximum log-degree cost.

The same tools can be used to study the fluctuations of other stochastic processes evolving on random graphs, including biased random walks (in discrete or continuous time), systems of many random walkers (independent or correlated), and interacting particle systems such as the exclusion or the zero-range process \cite{spohn1991}. Other observables can also be considered to target different graph properties, including the sum of degrees squared, which can be related to information and disease spreading via the epidemic threshold \cite{pastor2015}, indicator functions related to occupations \cite{angeletti2015}, jump-type observables involving two nodes linked by a transition \cite{chetrite2014}, as well as observables depending on in- and out-degrees, when considering directed networks. The tilted matrix can be constructed explicitly for these examples, but might be too large to be diagonalized explicitly. In this case, numerical methods based on cloning \cite{giardina2006,lecomte2007a,nemoto2016} or adaptive sampling \cite{nemoto2016,nemoto2017b,ferre2018} can be used to compute large deviation functions by direct simulations of random walks.

More work is required to confirm that the transitions reported in \cite{bacco2015} and here are ``genuine'' dynamical phase transitions that become sharp in the thermodynamic limit of infinite graphs. For this problem, we expect degree-degree correlations to play an important role, as the transitions are related to confined and low-connected regions of random graphs that lack the uncorrelated property of the whole \ER graph ensemble. Degree-degree correlations should also be important when considering other graph ensembles, such as small-world graphs, and real networks, making the calculation and approximation of large deviation functions more difficult. 

To conclude, we note that all the results reported here are ergodic-type results that assume that the random walk has enough time to cover the entire graph \cite{maier2017}. This means mathematically that the long-time limit $n\ra\infty$ is taken before the infinite-size limit $N\ra\infty$. The case where $n$ is allowed to scale with $N$, so as to study different regimes where, for example, the random walk has not yet covered the whole graph, is equally interesting but much more challenging to study.  

\begin{acknowledgments}
We thank Vito Latora, Vincenzo Nicosia, and Frank den Hollander for useful discussions. J.M.\ is supported by FCT, Portugal (Centre Grant UID/MULTI/04046/2013), while H.T.\ is supported by NRF, South Africa (Grant No.\ 90322 and No. 96199).
\end{acknowledgments}

\appendix
\section{Absorbing Markov chains}
\label{appsec}

\begin{figure}[b]
\centering
\includegraphics{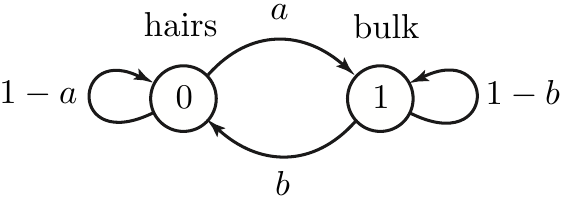}
\caption{Simple two-state Markov chain. For $b=0$, the state $1$ is absorbing.}
\label{fig7}
\end{figure}

Consider the Markov chain with states $\{0,1\}$ and transition matrix 
\be
\Pi = \left(
\begin{array}{cc}
1- a& a \\
b & 1-b
\end{array}
\right),
\ee
illustrated in Fig.~\ref{fig7}. For this example, it is easy to see that the rate function of the sample mean
\be
C_n = \frac{1}{n}\sum_{\ell=1}^n X_\ell,
\ee
where $X_\ell\in \{0,1\}$, becomes linear in $c$ as $b\ra 0$ \cite{whitelam2018}. In that limit, the state $1$ of the Markov chain becomes absorbing, which means that the probability of having $C_n = c$ is the survival probability $(1-a)^{n(1-c)}$ of staying in $0$ for $n(1-c)$ time steps, assuming that the random walk starts in $0$. Taking the large deviation limit \eqref{eqldt1}, we then find
\be
I(c) = (c-1) \ln (1-a),\qquad c\in [0,1].
\label{eqabrf1}
\ee 

\begin{figure}[t]
\centering
\includegraphics[width=2.7in]{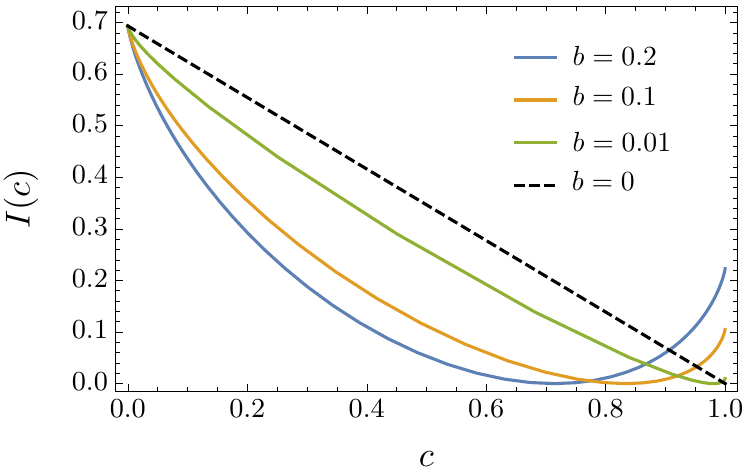}
\caption{(Color online) Rate function of the sample mean associated with the Markov chain of Fig.~\ref{fig7} for $a=1/2$ and different values of $b$. For $b=0$, $I(c)$ is a line.}
\label{fig8}
\end{figure}

This result is plotted in Fig.~\ref{fig8} and is compared with the rate functions obtained with the tilted matrix $\Pi_s$ for $0<b<1$, which become flat as $b\ra 0$. It can be checked by direct calculation that the derivative of the limiting SCGF jumps at $s=\ln (1-a)$, so there is a first-order DPT in the absorbing limit, accompanied by a closing gap between the two eigenvalues of the tilted matrix at the same critical value of $s$.

The analogy with the mean degree large deviations of the URW on ER graphs should be obvious. The state $0$ corresponds to the smallest degree fluctuation that can be achieved on a hair, whereas the state $1$ corresponds essentially to the average degree achieved on the bulk of the graph. Accordingly, $a$ represents the decay probability of going from a hair to the bulk, which is $1/2$ for ``long'' hairs with a node of degree 1 attached to a node of degree 2, while $b$ is the return probability of going from the bulk to a hair. In the thermodynamic limit, we expect the latter probability to decrease, since there are more edges within the bulk that keep the URW away from hairs. Whether $b$ vanishes in the thermodynamic limit determines, as mentioned, whether the  transition seen at the level of $\Psi'(s)$ in Figs.~\ref{fig1} and \ref{fig4} becomes a genuine DPT in that limit. 

\bibliography{masterbib}

\begin{thebibliography}{54}%
\makeatletter
\providecommand \@ifxundefined [1]{%
 \@ifx{#1\undefined}
}%
\providecommand \@ifnum [1]{%
 \ifnum #1\expandafter \@firstoftwo
 \else \expandafter \@secondoftwo
 \fi
}%
\providecommand \@ifx [1]{%
 \ifx #1\expandafter \@firstoftwo
 \else \expandafter \@secondoftwo
 \fi
}%
\providecommand \natexlab [1]{#1}%
\providecommand \enquote  [1]{``#1''}%
\providecommand \bibnamefont  [1]{#1}%
\providecommand \bibfnamefont [1]{#1}%
\providecommand \citenamefont [1]{#1}%
\providecommand \href@noop [0]{\@secondoftwo}%
\providecommand \href [0]{\begingroup \@sanitize@url \@href}%
\providecommand \@href[1]{\@@startlink{#1}\@@href}%
\providecommand \@@href[1]{\endgroup#1\@@endlink}%
\providecommand \@sanitize@url [0]{\catcode `\\12\catcode `\$12\catcode
  `\&12\catcode `\#12\catcode `\^12\catcode `\_12\catcode `\%12\relax}%
\providecommand \@@startlink[1]{}%
\providecommand \@@endlink[0]{}%
\providecommand \url  [0]{\begingroup\@sanitize@url \@url }%
\providecommand \@url [1]{\endgroup\@href {#1}{\urlprefix }}%
\providecommand \urlprefix  [0]{URL }%
\providecommand \Eprint [0]{\href }%
\providecommand \doibase [0]{http://dx.doi.org/}%
\providecommand \selectlanguage [0]{\@gobble}%
\providecommand \bibinfo  [0]{\@secondoftwo}%
\providecommand \bibfield  [0]{\@secondoftwo}%
\providecommand \translation [1]{[#1]}%
\providecommand \BibitemOpen [0]{}%
\providecommand \bibitemStop [0]{}%
\providecommand \bibitemNoStop [0]{.\EOS\space}%
\providecommand \EOS [0]{\spacefactor3000\relax}%
\providecommand \BibitemShut  [1]{\csname bibitem#1\endcsname}%
\let\auto@bib@innerbib\@empty
\bibitem [{\citenamefont {Barrat}\ \emph {et~al.}(2008)\citenamefont {Barrat},
  \citenamefont {Barth\'elemy},\ and\ \citenamefont {Vespignani}}]{barrat2008}%
  \BibitemOpen
  \bibfield  {author} {\bibinfo {author} {\bibfnamefont {A.}~\bibnamefont
  {Barrat}}, \bibinfo {author} {\bibfnamefont {M.}~\bibnamefont
  {Barth\'elemy}}, \ and\ \bibinfo {author} {\bibfnamefont {A.}~\bibnamefont
  {Vespignani}},\ }\href
  {http://www.cambridge.org/za/academic/subjects/physics/statistical-physics/dynamical-processes-complex-networks?format=HB&isbn=9780521879507#kG5uHEarO3wf2E6J.97}
  {\emph {\bibinfo {title} {Dynamical Processes on Complex Networks}}}\
  (\bibinfo  {publisher} {Cambridge University Press},\ \bibinfo {address}
  {Cambridge},\ \bibinfo {year} {2008})\BibitemShut {NoStop}%
\bibitem [{\citenamefont {Newman}(2010)}]{newman2010}%
  \BibitemOpen
  \bibfield  {author} {\bibinfo {author} {\bibfnamefont {M.~E.~J.}\
  \bibnamefont {Newman}},\ }\href
  {https://global.oup.com/academic/product/networks-9780199206650} {\emph
  {\bibinfo {title} {Networks: {A}n Introduction}}}\ (\bibinfo  {publisher}
  {Oxford University Press},\ \bibinfo {address} {Oxford},\ \bibinfo {year}
  {2010})\BibitemShut {NoStop}%
\bibitem [{\citenamefont {Pastor-Satorras}\ \emph {et~al.}(2015)\citenamefont
  {Pastor-Satorras}, \citenamefont {Castellano}, \citenamefont {Mieghem},\ and\
  \citenamefont {Vespignani}}]{pastor2015}%
  \BibitemOpen
  \bibfield  {author} {\bibinfo {author} {\bibfnamefont {R.}~\bibnamefont
  {Pastor-Satorras}}, \bibinfo {author} {\bibfnamefont {C.}~\bibnamefont
  {Castellano}}, \bibinfo {author} {\bibfnamefont {P.~Van}\ \bibnamefont
  {Mieghem}}, \ and\ \bibinfo {author} {\bibfnamefont {A.}~\bibnamefont
  {Vespignani}},\ }\bibfield  {title} {\enquote {\bibinfo {title} {Epidemic
  processes in complex networks},}\ }\href {\doibase 10.1103/RevModPhys.87.925}
  {\bibfield  {journal} {\bibinfo  {journal} {Rev. Mod. Phys.}\ }\textbf
  {\bibinfo {volume} {87}},\ \bibinfo {pages} {925--979} (\bibinfo {year}
  {2015})}\BibitemShut {NoStop}%
\bibitem [{\citenamefont {Kishore}\ \emph {et~al.}(2011)\citenamefont
  {Kishore}, \citenamefont {Santhanam},\ and\ \citenamefont
  {Amritkar}}]{kishore2011}%
  \BibitemOpen
  \bibfield  {author} {\bibinfo {author} {\bibfnamefont {V.}~\bibnamefont
  {Kishore}}, \bibinfo {author} {\bibfnamefont {M.~S.}\ \bibnamefont
  {Santhanam}}, \ and\ \bibinfo {author} {\bibfnamefont {R.~E.}\ \bibnamefont
  {Amritkar}},\ }\bibfield  {title} {\enquote {\bibinfo {title} {Extreme events
  on complex networks},}\ }\href {\doibase 10.1103/PhysRevLett.106.188701}
  {\bibfield  {journal} {\bibinfo  {journal} {Phys. Rev. Lett.}\ }\textbf
  {\bibinfo {volume} {106}},\ \bibinfo {pages} {188701} (\bibinfo {year}
  {2011})}\BibitemShut {NoStop}%
\bibitem [{\citenamefont {Kishore}\ \emph {et~al.}(2012)\citenamefont
  {Kishore}, \citenamefont {Santhanam},\ and\ \citenamefont
  {Amritkar}}]{kishore2012}%
  \BibitemOpen
  \bibfield  {author} {\bibinfo {author} {\bibfnamefont {V.}~\bibnamefont
  {Kishore}}, \bibinfo {author} {\bibfnamefont {M.~S.}\ \bibnamefont
  {Santhanam}}, \ and\ \bibinfo {author} {\bibfnamefont {R.~E.}\ \bibnamefont
  {Amritkar}},\ }\bibfield  {title} {\enquote {\bibinfo {title} {Extreme events
  and event size fluctuations in biased random walks on networks},}\ }\href
  {\doibase 10.1103/PhysRevE.85.056120} {\bibfield  {journal} {\bibinfo
  {journal} {Phys. Rev. E}\ }\textbf {\bibinfo {volume} {85}},\ \bibinfo
  {pages} {056120} (\bibinfo {year} {2012})}\BibitemShut {NoStop}%
\bibitem [{\citenamefont {Kishore}\ \emph {et~al.}(2013)\citenamefont
  {Kishore}, \citenamefont {Sonawane},\ and\ \citenamefont
  {Santhanam}}]{kishore2013}%
  \BibitemOpen
  \bibfield  {author} {\bibinfo {author} {\bibfnamefont {V.}~\bibnamefont
  {Kishore}}, \bibinfo {author} {\bibfnamefont {A.}~\bibnamefont {Sonawane}}, \
  and\ \bibinfo {author} {\bibfnamefont {M.~S.}\ \bibnamefont {Santhanam}},\
  }\bibfield  {title} {\enquote {\bibinfo {title} {Manipulation of extreme
  events on scale-free networks},}\ }\href {\doibase
  10.1103/PhysRevE.88.014801} {\bibfield  {journal} {\bibinfo  {journal} {Phys.
  Rev. E}\ }\textbf {\bibinfo {volume} {88}},\ \bibinfo {pages} {014801}
  (\bibinfo {year} {2013})}\BibitemShut {NoStop}%
\bibitem [{\citenamefont {Altarelli}\ \emph {et~al.}(2013)\citenamefont
  {Altarelli}, \citenamefont {Braunstein}, \citenamefont {L.~Dall'Asta},\ and\
  \citenamefont {Zecchina}}]{altarelli2013}%
  \BibitemOpen
  \bibfield  {author} {\bibinfo {author} {\bibfnamefont {F.}~\bibnamefont
  {Altarelli}}, \bibinfo {author} {\bibfnamefont {A.}~\bibnamefont
  {Braunstein}}, \bibinfo {author} {\bibfnamefont {L.}~\bibnamefont
  {L.~Dall'Asta}}, \ and\ \bibinfo {author} {\bibfnamefont {R.}~\bibnamefont
  {Zecchina}},\ }\bibfield  {title} {\enquote {\bibinfo {title} {Large
  deviations of cascade processes on graphs},}\ }\href {\doibase
  10.1103/PhysRevE.87.062115} {\bibfield  {journal} {\bibinfo  {journal} {Phys.
  Rev. E}\ }\textbf {\bibinfo {volume} {87}},\ \bibinfo {pages} {062115}
  (\bibinfo {year} {2013})}\BibitemShut {NoStop}%
\bibitem [{\citenamefont {Bianconi}(2017)}]{bianconi2017}%
  \BibitemOpen
  \bibfield  {author} {\bibinfo {author} {\bibfnamefont {G.}~\bibnamefont
  {Bianconi}},\ }\bibfield  {title} {\enquote {\bibinfo {title} {Fluctuations
  in percolation of sparse complex networks},}\ }\href {\doibase
  10.1103/PhysRevE.96.012302} {\bibfield  {journal} {\bibinfo  {journal} {Phys.
  Rev. E}\ }\textbf {\bibinfo {volume} {96}},\ \bibinfo {pages} {012302}
  (\bibinfo {year} {2017})}\BibitemShut {NoStop}%
\bibitem [{\citenamefont {Bianconi}(2018)}]{bianconi2018}%
  \BibitemOpen
  \bibfield  {author} {\bibinfo {author} {\bibfnamefont {G.}~\bibnamefont
  {Bianconi}},\ }\bibfield  {title} {\enquote {\bibinfo {title} {Rare events
  and discontinuous percolation transitions},}\ }\href {\doibase
  10.1103/PhysRevE.97.022314} {\bibfield  {journal} {\bibinfo  {journal} {Phys.
  Rev. E}\ }\textbf {\bibinfo {volume} {97}},\ \bibinfo {pages} {022314}
  (\bibinfo {year} {2018})}\BibitemShut {NoStop}%
\bibitem [{\citenamefont {Coghi}\ \emph {et~al.}(2018)\citenamefont {Coghi},
  \citenamefont {Radicchi},\ and\ \citenamefont {Bianconi}}]{coghi2018}%
  \BibitemOpen
  \bibfield  {author} {\bibinfo {author} {\bibfnamefont {F.}~\bibnamefont
  {Coghi}}, \bibinfo {author} {\bibfnamefont {F.}~\bibnamefont {Radicchi}}, \
  and\ \bibinfo {author} {\bibfnamefont {G.}~\bibnamefont {Bianconi}},\
  }\bibfield  {title} {\enquote {\bibinfo {title} {Controlling the uncertain
  response of real multiplex networks to random damage},}\ }\href {\doibase
  10.1103/PhysRevE.98.062317} {\bibfield  {journal} {\bibinfo  {journal} {Phys.
  Rev. E}\ }\textbf {\bibinfo {volume} {98}},\ \bibinfo {pages} {062317}
  (\bibinfo {year} {2018})}\BibitemShut {NoStop}%
\bibitem [{\citenamefont {Torrisi}\ \emph {et~al.}(2018)\citenamefont
  {Torrisi}, \citenamefont {Garetto},\ and\ \citenamefont
  {Leonardi}}]{torrisi2018}%
  \BibitemOpen
  \bibfield  {author} {\bibinfo {author} {\bibfnamefont {G.~L.}\ \bibnamefont
  {Torrisi}}, \bibinfo {author} {\bibfnamefont {M.}~\bibnamefont {Garetto}}, \
  and\ \bibinfo {author} {\bibfnamefont {E.}~\bibnamefont {Leonardi}},\
  }\bibfield  {title} {\enquote {\bibinfo {title} {A large deviation approach
  to super-critical bootstrap percolation on the random graph ${G}_{n,p}$},}\
  }\href {\doibase https://doi.org/10.1016/j.spa.2018.06.006} {\bibfield
  {journal} {\bibinfo  {journal} {Stoch. Proc. Appl.}\ } (\bibinfo {year}
  {2018}),\ https://doi.org/10.1016/j.spa.2018.06.006}\BibitemShut {NoStop}%
\bibitem [{\citenamefont {Dykman}\ \emph {et~al.}(2008)\citenamefont {Dykman},
  \citenamefont {Schwartz},\ and\ \citenamefont {Landsman}}]{dykman2008}%
  \BibitemOpen
  \bibfield  {author} {\bibinfo {author} {\bibfnamefont {M.~I.}\ \bibnamefont
  {Dykman}}, \bibinfo {author} {\bibfnamefont {I.~B.}\ \bibnamefont
  {Schwartz}}, \ and\ \bibinfo {author} {\bibfnamefont {A.~S.}\ \bibnamefont
  {Landsman}},\ }\bibfield  {title} {\enquote {\bibinfo {title} {Disease
  extinction in the presence of random vaccination},}\ }\href {\doibase
  10.1103/PhysRevLett.101.078101} {\bibfield  {journal} {\bibinfo  {journal}
  {Phys. Rev. Lett.}\ }\textbf {\bibinfo {volume} {101}},\ \bibinfo {pages}
  {078101} (\bibinfo {year} {2008})}\BibitemShut {NoStop}%
\bibitem [{\citenamefont {Lindley}\ \emph {et~al.}(2014)\citenamefont
  {Lindley}, \citenamefont {Shaw},\ and\ \citenamefont
  {Schwartz}}]{lindley2014}%
  \BibitemOpen
  \bibfield  {author} {\bibinfo {author} {\bibfnamefont {B.~S.}\ \bibnamefont
  {Lindley}}, \bibinfo {author} {\bibfnamefont {L.~B.}\ \bibnamefont {Shaw}}, \
  and\ \bibinfo {author} {\bibfnamefont {I.~B.}\ \bibnamefont {Schwartz}},\
  }\bibfield  {title} {\enquote {\bibinfo {title} {Rare-event extinction on
  stochastic networks},}\ }\href {\doibase 10.1209/0295-5075/108/58008}
  {\bibfield  {journal} {\bibinfo  {journal} {Europhys. Lett.}\ }\textbf
  {\bibinfo {volume} {108}},\ \bibinfo {pages} {58008} (\bibinfo {year}
  {2014})}\BibitemShut {NoStop}%
\bibitem [{\citenamefont {Hindes}\ and\ \citenamefont
  {Schwartz}(2016)}]{hindes2016}%
  \BibitemOpen
  \bibfield  {author} {\bibinfo {author} {\bibfnamefont {J.}~\bibnamefont
  {Hindes}}\ and\ \bibinfo {author} {\bibfnamefont {I.~B.}\ \bibnamefont
  {Schwartz}},\ }\bibfield  {title} {\enquote {\bibinfo {title} {Epidemic
  extinction and control in heterogeneous networks},}\ }\href {\doibase
  10.1103/PhysRevLett.117.028302} {\bibfield  {journal} {\bibinfo  {journal}
  {Phys. Rev. Lett.}\ }\textbf {\bibinfo {volume} {117}},\ \bibinfo {pages}
  {028302} (\bibinfo {year} {2016})}\BibitemShut {NoStop}%
\bibitem [{\citenamefont {Hindes}\ and\ \citenamefont
  {Schwartz}(2017{\natexlab{a}})}]{hindes2017}%
  \BibitemOpen
  \bibfield  {author} {\bibinfo {author} {\bibfnamefont {J.}~\bibnamefont
  {Hindes}}\ and\ \bibinfo {author} {\bibfnamefont {I.~B.}\ \bibnamefont
  {Schwartz}},\ }\bibfield  {title} {\enquote {\bibinfo {title} {Rare events in
  networks with internal and external noise},}\ }\href {\doibase
  10.1209/0295-5075/120/56004} {\bibfield  {journal} {\bibinfo  {journal}
  {Europhys. Lett.}\ }\textbf {\bibinfo {volume} {120}},\ \bibinfo {pages}
  {56004} (\bibinfo {year} {2017}{\natexlab{a}})}\BibitemShut {NoStop}%
\bibitem [{\citenamefont {Hindes}\ and\ \citenamefont
  {Schwartz}(2017{\natexlab{b}})}]{hindes2017b}%
  \BibitemOpen
  \bibfield  {author} {\bibinfo {author} {\bibfnamefont {J.}~\bibnamefont
  {Hindes}}\ and\ \bibinfo {author} {\bibfnamefont {I.~B.}\ \bibnamefont
  {Schwartz}},\ }\bibfield  {title} {\enquote {\bibinfo {title} {Epidemic
  extinction paths in complex networks},}\ }\href {\doibase
  10.1103/PhysRevE.95.052317} {\bibfield  {journal} {\bibinfo  {journal} {Phys.
  Rev. E}\ }\textbf {\bibinfo {volume} {95}},\ \bibinfo {pages} {052317}
  (\bibinfo {year} {2017}{\natexlab{b}})}\BibitemShut {NoStop}%
\bibitem [{\citenamefont {Hindes}\ and\ \citenamefont
  {Schwartz}(2017{\natexlab{c}})}]{hindes2017c}%
  \BibitemOpen
  \bibfield  {author} {\bibinfo {author} {\bibfnamefont {J.}~\bibnamefont
  {Hindes}}\ and\ \bibinfo {author} {\bibfnamefont {I.~B.}\ \bibnamefont
  {Schwartz}},\ }\bibfield  {title} {\enquote {\bibinfo {title} {Large order
  fluctuations, switching, and control in complex networks},}\ }\href {\doibase
  10.1038/s41598-017-08828-8} {\bibfield  {journal} {\bibinfo  {journal}
  {Scientific Reports}\ }\textbf {\bibinfo {volume} {7}},\ \bibinfo {pages}
  {10663} (\bibinfo {year} {2017}{\natexlab{c}})}\BibitemShut {NoStop}%
\bibitem [{\citenamefont {Bacco}\ \emph {et~al.}(2016)\citenamefont {Bacco},
  \citenamefont {Guggiola}, \citenamefont {K{\"u}hn},\ and\ \citenamefont
  {Paga}}]{bacco2015}%
  \BibitemOpen
  \bibfield  {author} {\bibinfo {author} {\bibfnamefont {C.~De}\ \bibnamefont
  {Bacco}}, \bibinfo {author} {\bibfnamefont {A.}~\bibnamefont {Guggiola}},
  \bibinfo {author} {\bibfnamefont {R.}~\bibnamefont {K{\"u}hn}}, \ and\
  \bibinfo {author} {\bibfnamefont {P.}~\bibnamefont {Paga}},\ }\bibfield
  {title} {\enquote {\bibinfo {title} {Rare events statistics of random walks
  on networks: {L}ocalization and other dynamical phase transitions},}\ }\href
  {\doibase 10.1088/1751-8113/49/18/184003} {\bibfield  {journal} {\bibinfo
  {journal} {J. Phys. A: Math. Theor.}\ }\textbf {\bibinfo {volume} {49}},\
  \bibinfo {pages} {184003} (\bibinfo {year} {2016})}\BibitemShut {NoStop}%
\bibitem [{\citenamefont {Dembo}\ and\ \citenamefont
  {Zeitouni}(1998)}]{dembo1998}%
  \BibitemOpen
  \bibfield  {author} {\bibinfo {author} {\bibfnamefont {A.}~\bibnamefont
  {Dembo}}\ and\ \bibinfo {author} {\bibfnamefont {O.}~\bibnamefont
  {Zeitouni}},\ }\href {http://www.springer.com/us/book/9783642033100} {\emph
  {\bibinfo {title} {Large Deviations Techniques and Applications}}},\ \bibinfo
  {edition} {2nd}\ ed.\ (\bibinfo  {publisher} {Springer},\ \bibinfo {address}
  {New York},\ \bibinfo {year} {1998})\BibitemShut {NoStop}%
\bibitem [{\citenamefont {{den Hollander}}(2000)}]{hollander2000}%
  \BibitemOpen
  \bibfield  {author} {\bibinfo {author} {\bibfnamefont {F.}~\bibnamefont {{den
  Hollander}}},\ }\href {http://bookstore.ams.org/fim-14.s} {\emph {\bibinfo
  {title} {Large Deviations}}},\ Fields Institute Monograph\ (\bibinfo
  {publisher} {AMS},\ \bibinfo {address} {Providence},\ \bibinfo {year}
  {2000})\BibitemShut {NoStop}%
\bibitem [{\citenamefont {Touchette}(2009)}]{touchette2009}%
  \BibitemOpen
  \bibfield  {author} {\bibinfo {author} {\bibfnamefont {H.}~\bibnamefont
  {Touchette}},\ }\bibfield  {title} {\enquote {\bibinfo {title} {The large
  deviation approach to statistical mechanics},}\ }\href {\doibase
  10.1016/j.physrep.2009.05.002} {\bibfield  {journal} {\bibinfo  {journal}
  {Phys. Rep.}\ }\textbf {\bibinfo {volume} {478}},\ \bibinfo {pages} {1--69}
  (\bibinfo {year} {2009})}\BibitemShut {NoStop}%
\bibitem [{\citenamefont {Jack}\ and\ \citenamefont
  {Sollich}(2010)}]{jack2010b}%
  \BibitemOpen
  \bibfield  {author} {\bibinfo {author} {\bibfnamefont {R.~L.}\ \bibnamefont
  {Jack}}\ and\ \bibinfo {author} {\bibfnamefont {P.}~\bibnamefont {Sollich}},\
  }\bibfield  {title} {\enquote {\bibinfo {title} {Large deviations and
  ensembles of trajectories in stochastic models},}\ }\href {\doibase
  10.1143/PTPS.184.304} {\bibfield  {journal} {\bibinfo  {journal} {Prog.
  Theoret. Phys. Suppl.}\ }\textbf {\bibinfo {volume} {184}},\ \bibinfo {pages}
  {304--317} (\bibinfo {year} {2010})}\BibitemShut {NoStop}%
\bibitem [{\citenamefont {Chetrite}\ and\ \citenamefont
  {Touchette}(2013)}]{chetrite2013}%
  \BibitemOpen
  \bibfield  {author} {\bibinfo {author} {\bibfnamefont {R.}~\bibnamefont
  {Chetrite}}\ and\ \bibinfo {author} {\bibfnamefont {H.}~\bibnamefont
  {Touchette}},\ }\bibfield  {title} {\enquote {\bibinfo {title}
  {Nonequilibrium microcanonical and canonical ensembles and their
  equivalence},}\ }\href {\doibase 10.1103/PhysRevLett.111.120601} {\bibfield
  {journal} {\bibinfo  {journal} {Phys. Rev. Lett.}\ }\textbf {\bibinfo
  {volume} {111}},\ \bibinfo {pages} {120601} (\bibinfo {year}
  {2013})}\BibitemShut {NoStop}%
\bibitem [{\citenamefont {Chetrite}\ and\ \citenamefont
  {Touchette}(2015{\natexlab{a}})}]{chetrite2014}%
  \BibitemOpen
  \bibfield  {author} {\bibinfo {author} {\bibfnamefont {R.}~\bibnamefont
  {Chetrite}}\ and\ \bibinfo {author} {\bibfnamefont {H.}~\bibnamefont
  {Touchette}},\ }\bibfield  {title} {\enquote {\bibinfo {title}
  {Nonequilibrium {M}arkov processes conditioned on large deviations},}\ }\href
  {\doibase 10.1007/s00023-014-0375-8} {\bibfield  {journal} {\bibinfo
  {journal} {Ann. Henri Poincar\'e}\ }\textbf {\bibinfo {volume} {16}},\
  \bibinfo {pages} {2005--2057} (\bibinfo {year}
  {2015}{\natexlab{a}})}\BibitemShut {NoStop}%
\bibitem [{\citenamefont {Chetrite}\ and\ \citenamefont
  {Touchette}(2015{\natexlab{b}})}]{chetrite2015}%
  \BibitemOpen
  \bibfield  {author} {\bibinfo {author} {\bibfnamefont {R.}~\bibnamefont
  {Chetrite}}\ and\ \bibinfo {author} {\bibfnamefont {H.}~\bibnamefont
  {Touchette}},\ }\bibfield  {title} {\enquote {\bibinfo {title} {Variational
  and optimal control representations of conditioned and driven processes},}\
  }\href {\doibase 10.1088/1742-5468/2015/12/P12001} {\bibfield  {journal}
  {\bibinfo  {journal} {J. Stat. Mech.}\ }\textbf {\bibinfo {volume} {2015}},\
  \bibinfo {pages} {P12001} (\bibinfo {year} {2015}{\natexlab{b}})}\BibitemShut
  {NoStop}%
\bibitem [{\citenamefont {Burda}\ \emph {et~al.}(2009)\citenamefont {Burda},
  \citenamefont {Duda}, \citenamefont {Luck},\ and\ \citenamefont
  {Waclaw}}]{burda2009}%
  \BibitemOpen
  \bibfield  {author} {\bibinfo {author} {\bibfnamefont {Z.}~\bibnamefont
  {Burda}}, \bibinfo {author} {\bibfnamefont {J.}~\bibnamefont {Duda}},
  \bibinfo {author} {\bibfnamefont {J.~M.}\ \bibnamefont {Luck}}, \ and\
  \bibinfo {author} {\bibfnamefont {B.}~\bibnamefont {Waclaw}},\ }\bibfield
  {title} {\enquote {\bibinfo {title} {Localization of the maximal entropy
  random walk},}\ }\href {\doibase 10.1103/PhysRevLett.102.160602} {\bibfield
  {journal} {\bibinfo  {journal} {Phys. Rev. Lett.}\ }\textbf {\bibinfo
  {volume} {102}},\ \bibinfo {pages} {160602} (\bibinfo {year}
  {2009})}\BibitemShut {NoStop}%
\bibitem [{\citenamefont {Nechaev}\ \emph {et~al.}(2017)\citenamefont
  {Nechaev}, \citenamefont {Tamm},\ and\ \citenamefont {Valba}}]{nechaev2017}%
  \BibitemOpen
  \bibfield  {author} {\bibinfo {author} {\bibfnamefont {S.~K.}\ \bibnamefont
  {Nechaev}}, \bibinfo {author} {\bibfnamefont {M.~V.}\ \bibnamefont {Tamm}}, \
  and\ \bibinfo {author} {\bibfnamefont {O.~V.}\ \bibnamefont {Valba}},\
  }\bibfield  {title} {\enquote {\bibinfo {title} {Path counting on simple
  graphs: {F}rom escape to localization},}\ }\href {\doibase
  10.1088/1742-5468/aa680a} {\bibfield  {journal} {\bibinfo  {journal} {J.
  Stat. Mech.}\ }\textbf {\bibinfo {volume} {2017}},\ \bibinfo {pages} {053301}
  (\bibinfo {year} {2017})}\BibitemShut {NoStop}%
\bibitem [{\citenamefont {G{\'o}mez-Gardenes}\ and\ \citenamefont
  {Latora}(2008)}]{gomez2008}%
  \BibitemOpen
  \bibfield  {author} {\bibinfo {author} {\bibfnamefont {J.}~\bibnamefont
  {G{\'o}mez-Gardenes}}\ and\ \bibinfo {author} {\bibfnamefont
  {V.}~\bibnamefont {Latora}},\ }\bibfield  {title} {\enquote {\bibinfo {title}
  {Entropy rate of diffusion processes on complex networks},}\ }\href {\doibase
  10.1103/PhysRevE.78.065102} {\bibfield  {journal} {\bibinfo  {journal} {Phys.
  Rev. E}\ }\textbf {\bibinfo {volume} {78}},\ \bibinfo {pages} {065102}
  (\bibinfo {year} {2008})}\BibitemShut {NoStop}%
\bibitem [{Note1()}]{Note1}%
  \BibitemOpen
  \bibinfo {note} {For finite, ergodic Markov chains, $\lambda (k)$ is actually
  analytic in $k$ \cite {dembo1998}. Care must be taken when considering the
  limit of infinite-size graphs, as well as graphs that are not all connected.
  This is discussed later in the text.}\BibitemShut {Stop}%
\bibitem [{Note2()}]{Note2}%
  \BibitemOpen
  \bibinfo {note} {The unicity of the root follows from the fact that $\Psi
  (s)$ is convex by definition and strictly convex for ergodic processes on a
  finite state-space \cite {dembo1998}.}\BibitemShut {Stop}%
\bibitem [{\citenamefont {Ole{\'s}}(2011)}]{oles2011}%
  \BibitemOpen
  \bibfield  {author} {\bibinfo {author} {\bibfnamefont {A.~K.}\ \bibnamefont
  {Ole{\'s}}},\ }\emph {\bibinfo {title} {Correlations in Random Graphs}},\
  \href
  {http://www.fais.uj.edu.pl/documents/41628/ee7e15df-3c10-4e2b-8547-100653fbbdd0}
  {Ph.D. thesis},\ \bibinfo  {school} {Jagiellonian University}, \bibinfo
  {address} {Krakow, Poland} (\bibinfo {year} {2011})\BibitemShut {NoStop}%
\bibitem [{\citenamefont {Tishby}\ \emph {et~al.}(2018)\citenamefont {Tishby},
  \citenamefont {Biham}, \citenamefont {Katzav},\ and\ \citenamefont
  {K\"uhn}}]{tishby2018}%
  \BibitemOpen
  \bibfield  {author} {\bibinfo {author} {\bibfnamefont {I.}~\bibnamefont
  {Tishby}}, \bibinfo {author} {\bibfnamefont {O.}~\bibnamefont {Biham}},
  \bibinfo {author} {\bibfnamefont {E.}~\bibnamefont {Katzav}}, \ and\ \bibinfo
  {author} {\bibfnamefont {R.}~\bibnamefont {K\"uhn}},\ }\bibfield  {title}
  {\enquote {\bibinfo {title} {Revealing the microstructure of the giant
  component in random graph ensembles},}\ }\href {\doibase
  10.1103/PhysRevE.97.042318} {\bibfield  {journal} {\bibinfo  {journal} {Phys.
  Rev. E}\ }\textbf {\bibinfo {volume} {97}},\ \bibinfo {pages} {042318}
  (\bibinfo {year} {2018})}\BibitemShut {NoStop}%
\bibitem [{Note3()}]{Note3}%
  \BibitemOpen
  \bibinfo {note} {This eigenvalue equation is different from the one
  considered in \cite {bacco2015} because of a different convention used for
  labeling the transition matrix elements, although the dominant eigenvalue
  giving the SCGF is the same.}\BibitemShut {Stop}%
\bibitem [{Note4()}]{Note4}%
  \BibitemOpen
  \bibinfo {note} {The averaging over graphs does not change the fact that
  $\Psi (s)$ is differentiable for finite graphs.}\BibitemShut {Stop}%
\bibitem [{Note5()}]{Note5}%
  \BibitemOpen
  \bibinfo {note} {We say ``nearly linearly'' because $\Psi '(s)$ does not have
  a jump singularity when $N<\infty $, which is a necessary condition for
  $I(c)$ or its convex envelope to have a linear part \cite
  {touchette2009}.}\BibitemShut {Stop}%
\bibitem [{Note6()}]{Note6}%
  \BibitemOpen
  \bibinfo {note} {To test this, we have calculated the dispersion (standard
  error over mean) associated with the distribution of $r_s(i)$ over all nodes
  $i$ having the same degree, to find that it does not decay with $N$. This was
  verified for different degrees, as well as different values of $\alpha $ and
  $s$.}\BibitemShut {Stop}%
\bibitem [{Note7()}]{Note7}%
  \BibitemOpen
  \bibinfo {note} {That term should also appear as an extra $k$ term in
  Eq.~(11) of \cite {bacco2015}.}\BibitemShut {Stop}%
\bibitem [{\citenamefont {Ephraim}\ and\ \citenamefont
  {Merhav}(2002)}]{ephraim2002}%
  \BibitemOpen
  \bibfield  {author} {\bibinfo {author} {\bibfnamefont {Y.}~\bibnamefont
  {Ephraim}}\ and\ \bibinfo {author} {\bibfnamefont {N.}~\bibnamefont
  {Merhav}},\ }\bibfield  {title} {\enquote {\bibinfo {title} {Hidden {M}arkov
  processes},}\ }\href {\doibase 10.1109/TIT.2002.1003838} {\bibfield
  {journal} {\bibinfo  {journal} {IEEE Trans. Info. Th.}\ }\textbf {\bibinfo
  {volume} {48}},\ \bibinfo {pages} {1518--1569} (\bibinfo {year}
  {2002})}\BibitemShut {NoStop}%
\bibitem [{\citenamefont {Rohwer}\ \emph {et~al.}(2015)\citenamefont {Rohwer},
  \citenamefont {Angeletti},\ and\ \citenamefont {Touchette}}]{rohwer2014}%
  \BibitemOpen
  \bibfield  {author} {\bibinfo {author} {\bibfnamefont {C.~M.}\ \bibnamefont
  {Rohwer}}, \bibinfo {author} {\bibfnamefont {F.}~\bibnamefont {Angeletti}}, \
  and\ \bibinfo {author} {\bibfnamefont {H.}~\bibnamefont {Touchette}},\
  }\bibfield  {title} {\enquote {\bibinfo {title} {Convergence of large
  deviation estimators},}\ }\href {\doibase 10.1103/PhysRevE.92.052104}
  {\bibfield  {journal} {\bibinfo  {journal} {Phys. Rev. E}\ }\textbf {\bibinfo
  {volume} {92}},\ \bibinfo {pages} {052104} (\bibinfo {year}
  {2015})}\BibitemShut {NoStop}%
\bibitem [{Note8()}]{Note8}%
  \BibitemOpen
  \bibinfo {note} {De Bacco \protect \emph {et al.}~\cite {bacco2015} define
  the IPR differently with what corresponds in our notation to the right
  eigenvector $r_s(i)$ raised to the power 4. Here, we follow the definition
  used in quantum mechanics, noting that $r_s(i)$ and $r_s(i)^2$ have no
  probabilistic interpretation.}\BibitemShut {Stop}%
\bibitem [{\citenamefont {{Tsobgni Nyawo}}\ and\ \citenamefont
  {Touchette}(2016)}]{tsobgni2016}%
  \BibitemOpen
  \bibfield  {author} {\bibinfo {author} {\bibfnamefont {P.}~\bibnamefont
  {{Tsobgni Nyawo}}}\ and\ \bibinfo {author} {\bibfnamefont {H.}~\bibnamefont
  {Touchette}},\ }\bibfield  {title} {\enquote {\bibinfo {title} {Large
  deviations of the current for driven periodic diffusions},}\ }\href {\doibase
  10.1103/PhysRevE.94.032101} {\bibfield  {journal} {\bibinfo  {journal} {Phys.
  Rev. E}\ }\textbf {\bibinfo {volume} {94}},\ \bibinfo {pages} {032101}
  (\bibinfo {year} {2016})}\BibitemShut {NoStop}%
\bibitem [{\citenamefont {H\'ebert-Dufresne}\ and\ \citenamefont
  {Allard}(2018)}]{dufresne2018}%
  \BibitemOpen
  \bibfield  {author} {\bibinfo {author} {\bibfnamefont {L.}~\bibnamefont
  {H\'ebert-Dufresne}}\ and\ \bibinfo {author} {\bibfnamefont {A.}~\bibnamefont
  {Allard}},\ }\bibfield  {title} {\enquote {\bibinfo {title} {Smeared phase
  transitions in percolation on real complex networks},}\ }\href@noop {} {\
  (\bibinfo {year} {2018})},\ \Eprint {http://arxiv.org/abs/arXiv:1810.00735}
  {arXiv:1810.00735} \BibitemShut {NoStop}%
\bibitem [{\citenamefont {Cover}\ and\ \citenamefont
  {Thomas}(1991)}]{cover1991}%
  \BibitemOpen
  \bibfield  {author} {\bibinfo {author} {\bibfnamefont {T.~M.}\ \bibnamefont
  {Cover}}\ and\ \bibinfo {author} {\bibfnamefont {J.~A.}\ \bibnamefont
  {Thomas}},\ }\href@noop {} {\emph {\bibinfo {title} {Elements of Information
  Theory}}}\ (\bibinfo  {publisher} {John Wiley},\ \bibinfo {address} {New
  York},\ \bibinfo {year} {1991})\BibitemShut {NoStop}%
\bibitem [{\citenamefont {Sinatra}\ \emph {et~al.}(2011)\citenamefont
  {Sinatra}, \citenamefont {G{\'o}mez-Gardenes}, \citenamefont {Lambiotte},
  \citenamefont {Nicosia},\ and\ \citenamefont {Latora}}]{sinatra2011}%
  \BibitemOpen
  \bibfield  {author} {\bibinfo {author} {\bibfnamefont {R.}~\bibnamefont
  {Sinatra}}, \bibinfo {author} {\bibfnamefont {J.}~\bibnamefont
  {G{\'o}mez-Gardenes}}, \bibinfo {author} {\bibfnamefont {R.}~\bibnamefont
  {Lambiotte}}, \bibinfo {author} {\bibfnamefont {V.}~\bibnamefont {Nicosia}},
  \ and\ \bibinfo {author} {\bibfnamefont {V.}~\bibnamefont {Latora}},\
  }\bibfield  {title} {\enquote {\bibinfo {title} {Maximal-entropy random walks
  in complex networks with limited information},}\ }\href {\doibase
  10.1103/PhysRevE.83.030103} {\bibfield  {journal} {\bibinfo  {journal} {Phys.
  Rev. E}\ }\textbf {\bibinfo {volume} {83}},\ \bibinfo {pages} {030103}
  (\bibinfo {year} {2011})}\BibitemShut {NoStop}%
\bibitem [{\citenamefont {Seifert}(2005)}]{seifert2005}%
  \BibitemOpen
  \bibfield  {author} {\bibinfo {author} {\bibfnamefont {U.}~\bibnamefont
  {Seifert}},\ }\bibfield  {title} {\enquote {\bibinfo {title} {Entropy
  production along a stochastic trajectory and an integral fluctuation
  theorem},}\ }\href {http://link.aps.org/abstract/PRL/v95/e040602} {\bibfield
  {journal} {\bibinfo  {journal} {Phys. Rev. Lett.}\ }\textbf {\bibinfo
  {volume} {95}},\ \bibinfo {eid} {040602} (\bibinfo {year}
  {2005})}\BibitemShut {NoStop}%
\bibitem [{\citenamefont {Spohn}(1991)}]{spohn1991}%
  \BibitemOpen
  \bibfield  {author} {\bibinfo {author} {\bibfnamefont {H.}~\bibnamefont
  {Spohn}},\ }\href@noop {} {\emph {\bibinfo {title} {Large Scale Dynamics of
  Interacting Particles}}}\ (\bibinfo  {publisher} {Springer Verlag},\ \bibinfo
  {address} {Berlin},\ \bibinfo {year} {1991})\BibitemShut {NoStop}%
\bibitem [{\citenamefont {Angeletti}\ and\ \citenamefont
  {Touchette}(2016)}]{angeletti2015}%
  \BibitemOpen
  \bibfield  {author} {\bibinfo {author} {\bibfnamefont {F.}~\bibnamefont
  {Angeletti}}\ and\ \bibinfo {author} {\bibfnamefont {H.}~\bibnamefont
  {Touchette}},\ }\bibfield  {title} {\enquote {\bibinfo {title} {Diffusions
  conditioned on occupation measures},}\ }\href {\doibase 10.1063/1.4941384}
  {\bibfield  {journal} {\bibinfo  {journal} {J. Math. Phys.}\ }\textbf
  {\bibinfo {volume} {57}},\ \bibinfo {pages} {023303} (\bibinfo {year}
  {2016})}\BibitemShut {NoStop}%
\bibitem [{\citenamefont {Giardina}\ \emph {et~al.}(2006)\citenamefont
  {Giardina}, \citenamefont {Kurchan},\ and\ \citenamefont
  {Peliti}}]{giardina2006}%
  \BibitemOpen
  \bibfield  {author} {\bibinfo {author} {\bibfnamefont {C.}~\bibnamefont
  {Giardina}}, \bibinfo {author} {\bibfnamefont {J.}~\bibnamefont {Kurchan}}, \
  and\ \bibinfo {author} {\bibfnamefont {L.}~\bibnamefont {Peliti}},\
  }\bibfield  {title} {\enquote {\bibinfo {title} {Direct evaluation of
  large-deviation functions},}\ }\href
  {http://link.aps.org/abstract/PRL/v96/e120603} {\bibfield  {journal}
  {\bibinfo  {journal} {Phys. Rev. Lett.}\ }\textbf {\bibinfo {volume} {96}},\
  \bibinfo {eid} {120603} (\bibinfo {year} {2006})}\BibitemShut {NoStop}%
\bibitem [{\citenamefont {Lecomte}\ and\ \citenamefont
  {Tailleur}(2007)}]{lecomte2007a}%
  \BibitemOpen
  \bibfield  {author} {\bibinfo {author} {\bibfnamefont {V.}~\bibnamefont
  {Lecomte}}\ and\ \bibinfo {author} {\bibfnamefont {J.}~\bibnamefont
  {Tailleur}},\ }\bibfield  {title} {\enquote {\bibinfo {title} {A numerical
  approach to large deviations in continuous time},}\ }\href {\doibase
  10.1088/1742-5468/2007/03/P03004} {\bibfield  {journal} {\bibinfo  {journal}
  {J. Stat. Mech.}\ }\textbf {\bibinfo {volume} {2007}},\ \bibinfo {pages}
  {P03004} (\bibinfo {year} {2007})}\BibitemShut {NoStop}%
\bibitem [{\citenamefont {Nemoto}\ \emph {et~al.}(2016)\citenamefont {Nemoto},
  \citenamefont {Bouchet}, \citenamefont {Jack},\ and\ \citenamefont
  {Lecomte}}]{nemoto2016}%
  \BibitemOpen
  \bibfield  {author} {\bibinfo {author} {\bibfnamefont {T.}~\bibnamefont
  {Nemoto}}, \bibinfo {author} {\bibfnamefont {F.}~\bibnamefont {Bouchet}},
  \bibinfo {author} {\bibfnamefont {R.~L.}\ \bibnamefont {Jack}}, \ and\
  \bibinfo {author} {\bibfnamefont {V.}~\bibnamefont {Lecomte}},\ }\bibfield
  {title} {\enquote {\bibinfo {title} {Population-dynamics method with a
  multicanonical feedback control},}\ }\href {\doibase
  10.1103/PhysRevE.93.062123} {\bibfield  {journal} {\bibinfo  {journal} {Phys.
  Rev. E}\ }\textbf {\bibinfo {volume} {93}},\ \bibinfo {pages} {062123}
  (\bibinfo {year} {2016})}\BibitemShut {NoStop}%
\bibitem [{\citenamefont {Nemoto}\ \emph {et~al.}(2017)\citenamefont {Nemoto},
  \citenamefont {Jack},\ and\ \citenamefont {Lecomte}}]{nemoto2017b}%
  \BibitemOpen
  \bibfield  {author} {\bibinfo {author} {\bibfnamefont {T.}~\bibnamefont
  {Nemoto}}, \bibinfo {author} {\bibfnamefont {R.~L.}\ \bibnamefont {Jack}}, \
  and\ \bibinfo {author} {\bibfnamefont {V.}~\bibnamefont {Lecomte}},\
  }\bibfield  {title} {\enquote {\bibinfo {title} {Finite-size scaling of a
  first-order dynamical phase transition: {A}daptive population dynamics and an
  effective model},}\ }\href {\doibase 10.1103/PhysRevLett.118.115702}
  {\bibfield  {journal} {\bibinfo  {journal} {Phys. Rev. Lett.}\ }\textbf
  {\bibinfo {volume} {118}},\ \bibinfo {pages} {115702} (\bibinfo {year}
  {2017})}\BibitemShut {NoStop}%
\bibitem [{\citenamefont {Ferr\'e}\ and\ \citenamefont
  {Touchette}(2018)}]{ferre2018}%
  \BibitemOpen
  \bibfield  {author} {\bibinfo {author} {\bibfnamefont {G.}~\bibnamefont
  {Ferr\'e}}\ and\ \bibinfo {author} {\bibfnamefont {H.}~\bibnamefont
  {Touchette}},\ }\bibfield  {title} {\enquote {\bibinfo {title} {Adaptive
  sampling of large deviations},}\ }\href {\doibase 10.1007/s10955-018-2108-8}
  {\bibfield  {journal} {\bibinfo  {journal} {J. Stat. Phys.}\ }\textbf
  {\bibinfo {volume} {172}},\ \bibinfo {pages} {1525--1544} (\bibinfo {year}
  {2018})}\BibitemShut {NoStop}%
\bibitem [{\citenamefont {Maier}\ and\ \citenamefont
  {Brockmann}(2017)}]{maier2017}%
  \BibitemOpen
  \bibfield  {author} {\bibinfo {author} {\bibfnamefont {B.~F.}\ \bibnamefont
  {Maier}}\ and\ \bibinfo {author} {\bibfnamefont {D.}~\bibnamefont
  {Brockmann}},\ }\bibfield  {title} {\enquote {\bibinfo {title} {Cover time
  for random walks on arbitrary complex networks},}\ }\href {\doibase
  10.1103/PhysRevE.96.042307} {\bibfield  {journal} {\bibinfo  {journal} {Phys.
  Rev. E}\ }\textbf {\bibinfo {volume} {96}},\ \bibinfo {pages} {042307}
  (\bibinfo {year} {2017})}\BibitemShut {NoStop}%
\bibitem [{\citenamefont {Whitelam}(2018)}]{whitelam2018}%
  \BibitemOpen
  \bibfield  {author} {\bibinfo {author} {\bibfnamefont {S.}~\bibnamefont
  {Whitelam}},\ }\bibfield  {title} {\enquote {\bibinfo {title} {Large
  deviations in the presence of cooperativity and slow dynamics},}\ }\href
  {\doibase 10.1103/PhysRevE.97.062109} {\bibfield  {journal} {\bibinfo
  {journal} {Phys. Rev. E}\ }\textbf {\bibinfo {volume} {97}},\ \bibinfo
  {pages} {062109} (\bibinfo {year} {2018})}\BibitemShut {NoStop}%
\end{thebibliography}%

\end{document}